\documentclass{aastex61}

\renewcommand{\(}{\begin{equation}}
\renewcommand{\)}{\end{equation}}
\newcommand{\bea}{\begin{eqnarray}}
\newcommand{\eea}{\end{eqnarray}}

\newcommand{\beq}{\begin{equation}}
\newcommand{\eeq}{\end{equation}}
\renewcommand{\(}{\begin{equation}}
\renewcommand{\)}{\end{equation}}
\def\cor#1{\langle #1 \rangle}
\def\bbf#1{{#1}}
\def\bbbf#1{{#1}}
\def\hsp{,\hspace{.2cm}}

\received{2016 Aug 31}
\revised{2017 May 6}
\accepted{2017 May 7}
\published{2017 May 30}
\submitjournal{ApJ}

\shorttitle{Quantifying Equilibrium}
\shortauthors{Evslin and Del Popolo}

\begin{document}

\title{Quantifying Departures from Equilibrium\\with the Spherical Jeans Equation}

\correspondingauthor{Jarah Evslin}
\email{jarah@impcas.ac.cn}

\author{Jarah Evslin}
\affiliation{Institute of Modern Physics, CAS, NanChangLu 509, Lanzhou 730000, China}
\affiliation{University of the Chinese Academy of Sciences, YuQuanLu 19A, Beijing 100049, China}

\author{Antonino Del Popolo}
\affiliation{Dipartimento di Fisica e Astronomia, University Of Catania, Viale Andrea Doria 6, 95125 Catania, Italy}
\affiliation{INFN sezione di Catania, Via S. Sofia 64, I-95123 Catania, Italy}
\affiliation{International Institute of Physics, Universidade Federal do Rio Grande do Norte, 59012-970 Natal, Brazil}


\begin{abstract}

Proper motions of collisionless pointlike objects in a spherically symmetric system, for example stars in a galaxy, can be used to test whether that system is in equilibrium, with no assumptions regarding isotropy.  In particular, the fourth order spherical Jeans equation yields expressions for two observable quantities characterizing the departure from equilibrium, both of which can be expressed in terms of time derivatives of first and third moments of the velocities.  As illustrations, we compute these quantities for tracer distributions drawn from an exact equilibrium configuration and also from near equilibrium configurations generated using the $N$-body code GALIC. 

\end{abstract}




\section{Introduction}
The $\Lambda$CDM standard cosmological model successfully reproduces a wealth of observations at large scales \citep{wmap,antonino14}.  At the kpc scale the confrontation of $\Lambda$CDM predictions with observations is more challenging.  For example, $\Lambda$CDM makes the robust prediction that pure dark matter halos will have cuspy profiles \citep{nfw}.  However in Nature no pure dark matter halo has yet been observed, and the dark matter halos hosting galaxies may be deformed by baryonic physics.  

Various studies disagree on just how much baryonic matter is necessary for such a deformation, although many studies \citep{governato, antoninojcap, onorbe}, but not all \citep{read}, find that ultrafaint dwarf (UFD) galaxies have little enough baryonic matter so that pure dark matter simulations should reliably reproduce their profiles.  However the dark matter profiles of UFDs are as of yet completely unknown.

In the case of the larger classical dwarf spheroidal (dSph) galaxies, various analyses of observations suggest cored \citep{goerdt,wp11} or cusped \citep{malcolm2014} density profiles.  Unfortunately $\Lambda$CDM simulations also differ in their predictions of whether these systems should have cusped or cored profiles, with the recent FIRE simulations \citep{onorbe} suggesting that baryonic physics generically leads to cored halos, while \citet{governato} suggests that only the heaviest are cored.  On the other hand Illustris simulations \citep{illustris} find that even the Milky Way is well fit by a cusped NFW profile, at least at the radii considered, and Eagle simulations \citep{eagle} find that baryonic physics leads to an even cuspier profiles in the case of massive galaxies.  Therefore even if observations some day are able to clearly distinguish a core from a cusp in classical dSphs, they will nonetheless not provide a foolproof test of $\Lambda$CDM.


In the next two decades, 30 meter class telescopes will revolutionize our understanding of the shapes of the dark matter halos of the Milky Way's satellite dSphs and UFDs by providing proper motions for hundreds to tens of thousands of members in each \citep{tmt}.  These telescopes will be able to cleanly distinguish an inner core from a cusp \citep{megaia} and may well be able to confront the ellipticities of the halos with simulations. Observation programs are being developed for these projects now, and for their planning one needs to know just which parameters will be determined.

Critical to all of these analyses is an understanding of the assumptions which go into the determination of the halo parameters.  To date, only line of sight velocities have been obtained and analyzed for these systems.  Three strong assumptions have been used so far in such analyses.  First, it has often been assumed that the systems are spherically symmetric {\bbf{\citep{wp11,malcolm2014}}} or at least axisymmetric {\bbf{\citep{1112.0319,1507.07620}}} despite the fact the CDM simulations predict triaxial halos \citep{frenkasse} and observations demonstrate that the stellar distribution is aspherical \citep{irwin}.  Second, often assumptions are made concerning the isotropy of the velocities.  In particular one often assumes that the anisotropy is independent of the radius {\bbf{\citep{0906.0341}}} or that the radial dependence is of a given functional form {\bbf{\citep{mk,1009.4857,1504.02048,1504.03309}}}.  Finally, it has been assumed that these systems are in equilibrium {\bbf{\citep{mk,malcolm2014}}}.   Eventually all three of these assumptions will need to be relaxed.  As a small step in this direction, in the present note {\bbf{we present a method which allows the third assumption to be tested given the proper motions of sufficiently many tracers.  This test is entirely independent of the second assumption, however it still depends on the first}}.  In the future, we intend to use the results presented below as a starting point for a treatment of the first assumption.   

{\bbf{The goal of this program is to provide the necessary tools to resolve the cusp/core debate using the precise astrometry which will be available on future extremely large telescopes.  This resolution can be achieved only if one is able to determine which systems are in equilibrium and also if and at what radii the equilibrium assumption breaks down.  More precisely}}  the establishment of equilibrium is necessary to turn these observations into a subhalo density profile, which in turn can teach one about the formation histories of the systems and about the nature of dark matter itself.  {\bbf{Given a dispersion-supported equilibrium configuration in a gravitational well, many formalisms exist to determine the underlying density profile which sources the gravitational potential.  Below we will consider the Jeans equations which work with the moments of the velocities.  However there are also orbit-based methods \citep{schwarz,rix} which incorporate the fact that the objects sourcing the potential must trace orbital solutions given the potential.}}

With proper motion data it will be possible to distinguish members {\bbf{of galaxies}} from contaminants at high radii, where foreground stars and background galaxies dominate.  There have been numerous reports of stellar excesses beyond the putative tidal radius {\bbf{\citep{0705.2901,1309.2799}, sometimes with expected nonequilbrium features \citep{1503.03896,1211.4875} but sometimes without \citep{hayashitidal}}}.   If it can be shown that they are in equilibrium, these distant stars can be used to probe a dark matter subhalo at large radii, providing a unique test of the $1/r^3$ density profile common to both cold and warm dark matter.  One may also test to see whether there is a gravitationally self-consistent solution to the subhalo profile given the tidal interactions with the host which may be inferred from the satellite galaxy proper motions.  If either of these tests is negative, it would provide evidence for a long range, nongravitational force.

{\bbf{The simplest way to test the equilibrium of a spherically symmetric system is to measure the odd moments of the stellar velocities.  In equilibrium these must vanish.  This provides a necessary but not sufficient condition for equilibrium.  In this paper we introduce a novel, complimentary necessary condition for equilibrium.}} We will use the second and fourth order spherical Jeans equations to determine the time derivatives of {\bbf{certain odd}} moments of stellar velocities {\bbf{in terms of the even moments}}.  A nonvanishing time derivative of any moment implies that the system is not in equilibrium, although the converse is not necessarily true.  We will make no assumptions regarding either equilibrium or the isotropy of the velocities, nor regarding the form of the gravitational potential, however we will make the rather poor assumption of spherical symmetry.  We will find that, even in this ideal setting, the second order Jeans equation in no way restricts the time derivatives of moments as a result of a degeneracy between the gravitational potential and the time derivative of the mean radial motion.  However the fourth order Jeans equation fixes two distinct combinations of time derivatives of velocity moments, and explicit formulae for these moments will be given below.

In Sec.~\ref{jeansez} we will use the second and fourth order Jeans equations to derive the formulas for the time derivatives as functions of velocity moments in spherical coordinates.  Then in Sec.~\ref{projsez} we will provide an explicit map which yields the spherical coordinate moments given the observed projected 3-dimensional velocity moments.  {\bbbf{Next}} in Sec.~\ref{exsez} we will use the $N$-body code GALIC \citep{galic} to generate an example which both illustrates the results of Secs.~\ref{jeansez} and \ref{projsez} and also {\bbf{provides an estimate of the precision with which these time derivates may be determined}}. {\bbbf{In Sec.~\ref{tmtsez} we describe future observations which may allow such an analysis.  Concluding remarks appear in Sec.~\ref{consez}.}}


\section{Jeans analysis} \label{jeansez}

Let $f(x,v)$ be the phase-space distribution function of stars, in other words the phase-space density at a position $x$ and velocity $v$.  By Liouville's theorem \citep{b68,gibbs}, in a collisionless system $f(x,v)$ is constant along stellar trajectories
\beq
\frac{df}{dt}=0.
\eeq
One obtains Boltzmann's equation \citep{boltzmann} by expanding the total differential in terms of partial differentials via the chain rule
\bea
0=\frac{df}{dt}&=&\frac{\partial f}{\partial t}+v_r \frac{\partial f}{\partial r} + \left(\frac{v_\theta^2+v_\phi^2}{r}-\frac{d\Phi}{dr}\right)\frac{\partial f}{\partial v_r}\nonumber\\
&&+\frac{1}{r}\left(v_\phi^2{\rm{cot}}(\theta)-v_rv_\theta\right)\frac{\partial f}{\partial v_\theta}\nonumber\\
&&-\frac{1}{r}\left(v_\phi v_\theta{\rm{cot}}(\theta)+v_rv_\phi\right)\frac{\partial f}{\partial v_\phi} \label{boltzeq}
\eea
where $\Phi(r)$ is the gravitational potential.  The gravitational potential and the distribution $f(x,v)$ are both assumed to be spherically symmetric in the spherical coordinate system $(r,\theta,\phi)$ centered at some point in space.

The first term on the right hand side of Eq.~(\ref{boltzeq}) vanishes for a system in equilibrium.  Our analysis differs from many in the literature in that we do not assume that this term is zero, it is in no way constrained in our analysis.  Following \citet{mk} but keeping this extra term, we obtain the second order Jeans equation \citep{jeans} by multiplying by $v_r$ and integrating over the full 3-dimensional space of velocities
\beq
0=\frac{\partial}{\partial t}\left(\nu \langle v_r\rangle\right)+\frac{\partial}{\partial r}\left(\nu \langle v_r^2\rangle\right)+\frac{2\nu(r)}{r}\left(\langle v_r^2\rangle-\langle v_\theta^2\rangle\right)+\nu\frac{d\Phi}{dr} \label{secondo}
\eeq
where the stellar (position) density is
\beq
\nu(x)=\int f(x,v) d^3 v.
\eeq
The first term in Eq.~(\ref{secondo}) vanishes for a system in equilibrium.  It characterizes the radial acceleration of the system.  Note that observations can only determine $\langle v_r^2\rangle$ and $\langle v_\theta^2\rangle$, not $\Phi$, and so observationally it is not possible to separate the nonequilibrium and gravitational terms in (\ref{secondo}), at best one may determine their sum
\beq
\frac{\partial}{\partial t}\left(\nu \langle v_r\rangle\right)+\nu\frac{d\Phi}{dr}=-\frac{\partial}{\partial r}\left(\nu \langle v_r^2\rangle\right)-\frac{2\nu}{r}\left(\langle v_r^2\rangle-\langle v_\theta^2\rangle\right).
\eeq
As a result, it is not possible to test the equilibrium hypothesis using the second order spherical Jeans equation alone.

Similarly, the two fourth order spherical Jeans equations are obtained by multiplying the Boltzmann equation by $v_r^3$ and $v_r v_\theta^2$ respectively and integrating over velocities \citep{mk}
\beq
0  = \frac{\partial \left(\nu\langle v_r^3\rangle\right)}{\partial t} + \frac{\partial \left(\nu\langle v_r^4\rangle\right)}{\partial r}+\frac{2\nu \langle v_r^4\rangle}{r} - \frac{6\nu \langle v_r^2 v_\theta^2\rangle}{r} + 3\nu\langle v_r^2\rangle\frac{d\Phi}{dr} \label{quartoa}
\eeq
and
\bea
0&=&\frac{\partial \left(\nu\langle v_r v_\theta^2\rangle\right)}{\partial t}+\frac{\partial \left(\nu\langle v_r^2v_\theta^2\rangle\right)}{\partial r}\nonumber\\&&+\frac{4\nu \langle v_r^2v_\theta^2\rangle}{r}-\frac{4\nu \langle v_\theta^4\rangle}{3r}+\nu\langle v_\theta^2\rangle\frac{d\Phi}{dr} \label{quartob}
\eea
where we have used the identities
\beq
\cor{v_\theta^4}=\cor{v_\phi^4}=3\cor{v_\theta^2v_\phi^2} \label{ideq}
\eeq
which are consequences of spherical symmetry.

Again each equation contains both a nonequilibrium term and a gravitational potential term, and so once the velocity moments have been measured at a fixed time the fourth order equations consist of three unknowns but only two equations.  Including the second order equation, there is an additional unknown, the time derivative of $\cor{v_r}$, and so three equations and four unknowns.  However, it is possible to determine linear combinations of the unknown time derivatives, which vanish in an equilibrium system and so can be used to test the equilibrium hypothesis.

One simple way to combine the time derivatives into an expression that can be determined observationally is simply to use the second order Jeans equation to eliminate $d\Phi/dr$ in the fourth order equations.  The fourth order equations then become
\bea
&&\hspace{-.8cm}\frac{\partial \left(\nu\langle v_r^3\rangle\right)}{\partial t} -3\cor{v_r^2}\frac{\partial \left(\nu \langle v_r\rangle\right)}{\partial t} = -\frac{\partial \left(\nu\langle v_r^4\rangle\right)}{\partial r} + 3\cor{v_r^2}\frac{\partial\left(\nu \langle v_r^2\rangle\right)}{\partial r}\nonumber\\&&\hspace{-.1cm}-\frac{2\nu \left(\langle v_r^4\rangle-3\cor{v_r^2}^2\right)}{r}
+\frac{6\nu \left(\langle v_r^2 v_\theta^2\rangle-\cor{v_r^2}\cor{v_\theta^2}\right)}{r} 
\eea
and
\bea
&&\hspace{-.5cm}\frac{\partial \left(\nu\langle v_r v_\theta^2\rangle\right)}{\partial t}-\cor{v_\theta^2}\frac{\partial \left(\nu \langle v_r\rangle\right)}{\partial t}=-\frac{\partial \left(\nu\langle v_r^2v_\theta^2\rangle\right)}{\partial r}\nonumber\\&&+\cor{v_\theta^2}\frac{\partial}{\partial r}\left(\nu \langle v_r^2\rangle\right)+\frac{2\nu \left(-2\langle v_r^2v_\theta^2\rangle+\cor{v_r^2}\cor{v_\theta^2}\right)}{r}\nonumber\\&&+\frac{\nu \left(\frac{4}{3}\cor{v_\theta^4}-2\cor{v_\theta^2}^2\right)}{r} .
\eea
Now the left hand side of each equation characterizes the departure from equilibrium, and is equal to zero for an equilibrium configuration.  The right hand side consists of moments of instantaneous velocities, and so can in principle be extracted from observations of the 3-dimensional velocities, if projection effects can be removed.

In simulations (see Sec.~\ref{exsez}) we have found that the imprecision in this method is dominated by statistical fluctuations in the velocities.  These statistical fluctuations can be reduced if instead of $v_\theta$ one uses the tangential velocity $v_T$, defined to be the square root sum in quadrature of $v_\theta$ and $v_\phi$.  {\bbf{The reason for this is that, due to spherical symmetry, $v_\theta$ and $v_\phi$ are distributed identically.  As a result the total uncertainty on the sum $v^2_T=v^2_\theta+v^2_\phi$ is $\sqrt{2}$ times the uncertainty on $v^2_\theta$ or $v^2_\phi$.  Therefore the expected {\it{fractional}} fluctuations expected for the quantity $v_T^2$ are suppressed with respect to those of $v^2_\theta$ or $v^2_\phi$ by a factor of $\sqrt{2}$.  A similar argument shows that the fluctuations of the fourth moments are suppressed even more strongly when using $v_T$.}} 

In this case the above equations can be written
\bea
& &\hspace{-.8cm}\frac{\partial \left(\nu\langle v_r^3\rangle\right)}{\partial t} - 3\cor{v_r^2}\frac{\partial \left(\nu \langle v_r\rangle\right)}{\partial t} = -\frac{\partial \left(\nu\langle v_r^4\rangle\right)}{\partial r} +  3\cor{v_r^2}\frac{\partial\left(\nu \langle v_r^2\rangle\right)}{\partial r}\nonumber\\&&\hspace{-.1cm}-\frac{2\nu \left(\langle v_r^4\rangle-3\cor{v_r^2}^2\right)}{r}
+\frac{3\nu \left(\langle v_r^2 v_T^2\rangle-\cor{v_r^2}\cor{v_T^2}\right)}{r} \label{maina}
\eea
and
\bea
&&\hspace{-.5cm}\frac{\partial \left(\nu\langle v_r v_T^2\rangle \right)}{\partial t}-\cor{v_T^2}\frac{\partial \left(\nu \langle v_r\rangle\right)}{\partial t}=-\frac{\partial \left(\nu\langle v_r^2v_T^2\rangle \right)}{\partial r}\nonumber\\&&+\cor{v_T^2}\frac{\partial}{\partial r}\left(\nu \langle v_r^2\rangle\right)+\frac{2\nu \left(-2\langle v_r^2v_T^2\rangle+\cor{v_r^2}\cor{v_T^2} \right)}{r}\nonumber\\&&+\frac{\nu \left(\cor{v_T^4}-\cor{v_T^2}^2\right)}{r} . \label{mainb}
\eea

\section{Deprojection} \label{projsez}

Due to the projection of stars onto the celestial sphere, one cannot determine the spherical coordinates $(r,\theta,\phi)$ at a point, rather one measures the {\bbf{cylindrical}} coordinates $(Z,R,\Psi)$ where $Z$ is the line of sight direction, $R$ is the orthogonal to the line of sight and $\Psi$ is the azimuthal angle.  In practice $Z$ cannot be constrained with a precision which would be useful for our analysis.  Velocities in this {\bbf{cylindrical}} coordinate system are related to those in spherical coordinates via 
\bea
&&v_Z=\frac{\sqrt{r^2-R^2}}{r}v_r+\frac{R}{r}v_\theta\\
&&v_R= \frac{R}{r}v_r+\frac{\sqrt{r^2-R^2}}{r}v_\theta\hsp
v_\Psi=v_\phi.\nonumber
\eea

Integrating over the line of sight one recovers the standard formulas for the observed second velocity moments
\bea
\cor{v_Z^2}      &=&    \frac{2}{\eta}\int_R^\infty \frac{r\nu dr}{\sqrt{r^2-R^2}}\left(\left(1-\frac{R^2}{r^2}\right)\cor{v_r^2}+\frac{R^2}{r^2}\cor{v_\theta^2}\right)\nonumber\\
\cor{v_R^2}      &=&    \frac{2}{\eta}\int_R^\infty \frac{r\nu dr}{\sqrt{r^2-R^2}}\left(\frac{R^2}{r^2}\cor{v_r^2}+\left(1-\frac{R^2}{r^2}\right)\cor{v_\theta^2}\right)\nonumber\\
\cor{v_\Psi^2}      &=&    \frac{2}{\eta}\int_R^\infty \frac{r\nu dr}{\sqrt{r^2-R^2}}\cor{v_\phi^2} \label{dueeq}
\eea
where $\eta$ is the observed two dimensional density.
The 3-dimensional density $\nu$ at $r$ can be obtained from the observed 2-dimensional density $\eta$ via an inverse Abel transform 
\beq
\nu=-\frac{1}{\pi}\int_r^\infty \frac{dR}{\sqrt{R^2-r^2}}\frac{d\eta}{dR}. \label{luminv}
\eeq
Similarly an inverse Abel transform of the last equation in (\ref{dueeq}) yields the second moment
\beq
\cor{v_\theta^2}=\cor{v_\phi^2}=-\frac{1}{\pi\nu}\int_r^\infty \frac{dR}{\sqrt{R^2-r^2}}\frac{d\left(\eta\cor{v_\Psi^2}\right)}{dR}. \label{vt}
\eeq
Transforming the sum of the first two equations in (\ref{dueeq}) one finds
\beq
\cor{v_r^2}+\cor{v_\theta^2}=-\frac{1}{\pi\nu}\int_r^\infty \frac{dR}{\sqrt{R^2-r^2}}\frac{d\left(\eta\cor{v_Z^2}+\eta\cor{v_R^2}\right)}{dR}.
\eeq
Subtracting (\ref{vt}) one finds the last second moment
\beq
\cor{v_r^2}=-\frac{1}{\pi\nu}\int_r^\infty \frac{dR}{\sqrt{R^2-r^2}}\frac{d\left(\eta\left(\cor{v_Z^2}+\cor{v_R^2}-\cor{v_\Psi^2}\right)\right)}{dR}. \label{vr}
\eeq
Summarizing so far, Eqs.~(\ref{luminv}), (\ref{vt}) and (\ref{vr}) provide the 3-dimensional density and second moments, the expressions appearing in the second order Jeans equation (\ref{secondo}), from the observed projected density and second moments.  

{\bbf{The above equations express three unknowns {\bbf{$\cor{v_r^2},\ \cor{v_\theta^2}$\ and $\nu$}} in terms of four observables {\bbf{$\cor{v_R^2},\ \cor{v_Z^2},\ \cor{v_\Psi^2}$ and $\eta$}} using only three equations.   The fact that there is one more observable than equations implies that this system contains one exact degeneracy.  More precisely, there is a one-parameter family of {\bbbf{possible observations}} for which these equations yield the same set of unknowns.   As a consequence it would be possible to modify the set of three equations by shifting the observables along the degenerate direction without affecting the obtained estimates of the unknowns.  These equations are therefore not unique.  Such a shift of the equations along the degenerate direction would not affect their validity or physical content.  However the precisions with which the observables are measured do depend on the shift, and so such a shift would affect the resulting uncertainties in the estimates of the unknowns.  In summary, while the equations above are correct, they do not offer optimal precision in the determination of the unknowns.  This could be achieved, for example, with a $\chi^2$ fit of the observables to the line of sight projections of the unknowns.}}  

{\bbbf{In summary, the above set of equations is not unique and does not minimize the uncertainties achieved on the parameters.  We will nonetheless use them because they provide a fast estimation of the deprojected parameters, and computational speed will be important given the large number of simulations which we will analyze.  However, the deprojection of real data can, at the expense of greater computation time, be performed more precisely using standard methods such as a $\chi^2$ fit.}}

Similarly, the projected fourth moments are integrals along the line of sight of the spherical coordinate fourth moments
\bea
&&\hspace{-0.7cm}\cor{v_Z^4}=\frac{2}{\eta}\int_R^\infty \frac{r\nu dr}{\sqrt{r^2-R^2}}\label{quattroeq}\\
&&\hspace{-0.7cm}\times\left(\left(1-\frac{R^2}{r^2}\right)^2\cor{v_r^4}+6\left(\frac{R^2}{r^2}-\frac{R^4}{r^4}\right)\cor{v_r^2v_\theta^2}+\frac{R^4}{r^4}\cor{v_\theta^4}\right)\nonumber\\
&&\hspace{-0.7cm}\cor{v_R^4}=\frac{2}{\eta}\int_R^\infty \frac{r\nu dr}{\sqrt{r^2-R^2}}\nonumber\\
&&\hspace{-0.7cm}\times\left(\frac{R^4}{r^4}\cor{v_r^4}+6\left(\frac{R^2}{r^2}-\frac{R^4}{r^4}\right)\cor{v_r^2v_\theta^2}+\left(1-\frac{R^2}{r^2}\right)^2\cor{v_\theta^4}\right)\nonumber\\
&&\hspace{-0.7cm}\cor{v_\Psi^4}=\frac{2}{\eta}\int_R^\infty \frac{r\nu dr}{\sqrt{r^2-R^2}}\cor{v_\theta^4}\nonumber\\
&&\hspace{-0.7cm}\cor{v_Z^2v_R^2}=\frac{2}{\eta}\int_R^\infty \frac{r\nu dr}{\sqrt{r^2-R^2}}\nonumber\\
&&\hspace{-0.7cm}\times\left(\left(\frac{R^2}{r^2}  -  \frac{R^4}{r^4}\right)  \left(\cor{v_r^4}+\cor{v_\theta^4}\right)  +  \left(  1  -  6\frac{R^2}{r^2}  +6\frac{R^4}{r^4}\right)\cor{v_r^2v_\theta^2}\right)\nonumber\\
&&\hspace{-0.7cm}\cor{v_Z^2v_\Psi^2}=\frac{2}{\eta}\int_R^\infty   \frac{r\nu dr}{\sqrt{r^2-R^2}}
\left(\left(1-\frac{R^2}{r^2}\right)  \cor{v_r^2v_\theta^2}  +  \frac{R^2}{3r^2}\cor{v_\theta^4}\right)\nonumber\\
&&\hspace{-0.7cm}\cor{v_R^2v_\Psi^2}=\frac{2}{\eta}\int_R^\infty   \frac{r\nu dr}{\sqrt{r^2-R^2}}
\left(\frac{R^2}{r^2}\cor{v_r^2v_\theta^2}+\frac{1}{3}\left(1-\frac{R^2}{r^2}\right)\cor{v_\theta^4}\right).\nonumber
\eea
As in the case of second moments, one may obtain a combination in a form which is easily Abel-transformed by noting the identities
\beq
v_\Psi=v_\phi\hsp v_Z^2+v_R^2=v_r^2+v_\theta^2
\eeq
and using the three combinations of moments which are products of these two invariants
\bea
&&\hspace{-0.7cm}\cor{(v_Z^2+v_R^2)^2}=\frac{2}{\eta}\int_R^\infty \frac{r\nu dr}{\sqrt{r^2-R^2}}\cor{\left(v_r^2+v_\theta^2\right)^2}\nonumber\\
&&\hspace{-0.7cm}\cor{(v_Z^2+v_R^2)v_\Psi^2}=\frac{2}{\eta}\int_R^\infty \frac{r\nu dr}{\sqrt{r^2-R^2}}\left(\cor{v_r^2v_\theta^2}+\frac{1}{3}\cor{v_\theta^4}\right)\nonumber\\
&&\hspace{-0.7cm}\cor{v_\Psi^4}=\frac{2}{\eta}\int_R^\infty \frac{r\nu dr}{\sqrt{r^2-R^2}}\cor{v_\theta^4}.
\eea
Linear combinations of the inverse Abel transforms then yield the spherical coordinate moments in terms of the observed projected moments
\bea
&&\cor{v_r^4}=-\frac{1}{\pi\nu}\int_r^\infty \frac{dR}{\sqrt{R^2-r^2}}\nonumber\\
&&\times\frac{d}{dR}\left(\eta\left(\cor{\left(v_Z^2+v_R^2-v_\Psi^2\right)^2}-\frac{4}{3}\cor{v_\Psi^4}\right)\right)\nonumber\\
&&\cor{v_r^2v_\theta^2}=-\frac{1}{\pi\nu}\int_r^\infty \frac{dR}{\sqrt{R^2-r^2}}\nonumber\\
&&\times\frac{d}{dR}\left(\eta\left(\cor{\left(v_Z^2+v_R^2\right)v_\Psi^2}-\frac{1}{3}\cor{v_\Psi^4}\right)\right)\nonumber\\
&&\cor{v_\theta^4}=-\frac{1}{\pi\nu}\int_r^\infty \frac{dR}{\sqrt{R^2-r^2}}\frac{d\left(\eta\cor{v_\Psi^4}\right)}{dR}.  \label{quadabel}
\eea
{\bbf{The above equations express three unknowns $\cor{v_r^4},\ \cor{v_r^2v_\theta^2}$\ and $\cor{v_\theta^4}$ in terms of $\nu$ and seven observables $\cor{v_Z^4}, \cor{v_R^4}, \cor{v_Z^2v_R^2}, \cor{v_\Psi^4}, \ \cor{v_Z^2v_\Psi^2}$, $\cor{v_R^2v_\psi^2}$ and $\eta$ using only three equations.  As there are more observables than equations, again the obtained unknowns are invariant under transformations of the observables.  Again this means that the equations may be transformed into a form which reduces the resulting uncertainties, and so they are suboptimal.}}

In summary, one first observes the projected, $r$-dependent 2-dimensional density {\bbf{$\eta$}}, and the second and fourth order moments.  These can be transformed into spherical coordinates via the inverse Abel transforms (\ref{luminv}), (\ref{vt}), (\ref{vr}) and (\ref{quadabel}).  Finally, these spherical-coordinate fixed-time moments can be substituted into Eqs.~(\ref{maina}) and (\ref{mainb}) to obtain an estimate of the time-derivatives of the odd order moments.  These time derivatives are nonzero only if the system is not in equilibrium.  Thus, given spherical symmetry and sufficiently many proper motions\footnote{{\bbf{In the example below we will find that $10^4$ proper motions are sufficient for a nontrivial test of the equilibrium.}}}, one can test the equilibrium of the system without knowledge of the potential or any assumptions regarding the isotropy of the orbits.  This is our main result.

\section{Near Equilibrium Configurations from GALIC} \label{exsez}

\subsection{The Setup}

In this section we will present a simple example of Eqs.~(\ref{maina}) and (\ref{mainb}) at work.  We will use the $N$-body code GALIC \citep{galic} to produce systems of $10^5$ particles {\bbf{of mass $1.86\times 10^4\ \rm{M}_\odot$}} which interact gravitationally.  {\bbf{As the only interaction is gravitational, at this point each particle could be either dark or stellar matter.}}  We will consider both spherical and axisymmetric mass distributions, so that we may estimate the systematic errors created in our approach as a result of our false assumption of spherical symmetry.

These correspond to Hernquist dark matter halos \citep{hernquist}, with no other components, described by the GALIC standard models H1 and two variations of the model H6, each rescaled by a factor of ten to very roughly yield typical dwarf spheroidal galaxy dimensions.  In particular, the Virial radius $r_{200}$ of each halo is 20 kpc and the Hernquist parameter is $a=3.45$ kpc.  

{\bbf{For now}} we will assume that the true radial coordinates of the particles are known.  In reality, this requires parallax measurements with a precision unlikely to be obtained for stars in dwarf spheroidal galaxies in the foreseeable future.  As a result, we only test Sec.~\ref{jeansez}, not Sec.~\ref{projsez}.  In Subsec.~\ref{projsub} we will use the projected observables {\bbf{together with the deprojection in Sec.~\ref{projsez} to provide a comprehensive test of our results}}.

These made-to-measure simulations fix the density distribution of the particles and adjust the velocities towards an equilibrium configuration one at time, beginning with Gaussian density profiles that approximate those of Hernquist profiles.  {\bbf{Unfortunately such initial configurations are already quite close to equilibrium, and so this approach will not allow us to test our formalism far from equilibrium\footnote{A system far from equilibrium will be considered in Subsec.~\ref{hernsez}.}.}} Each time that a given particle is adjusted, the system is driven closer to, or at least not further from, equilibrium.  Thus by measuring the system after different numbers of iterations per particle, one can obtain velocity profiles for various distances from equilibrium.  

The arguments in Sec.~\ref{jeansez} imply that, as the systems approach equilibrium, the right hand sides of Eqs.~(\ref{maina}) and (\ref{mainb}) should approach zero for a spherically symmetric density profile, whereas one expects that a breakdown of spherical symmetry will lead to a nonzero limiting value. {\bbf{This is because spherical symmetry was used in the derivation of Eqs.~(\ref{secondo},\ref{quartoa},\ref{quartob}).  In the absence of spherical symmetry, in general the left hand side of each equation will be an arbitrary function of the position.  For a spherically symmetric equilibrium system, the time derivatives in these equations vanish and the right hand sides of Eqs.~(\ref{maina}) and (\ref{mainb}) are linear combinations of the left hand sides of Eqs.~(\ref{secondo},\ref{quartoa},\ref{quartob}), and so are also arbitrary functions of position.}}

To test this prediction, we let $\alpha_i$ and $\beta_i$ denote the four terms of the right hand sides of Eqs.~(\ref{maina}) and (\ref{mainb}) respectively, so that these equations imply that for an equilibrium configuration
\beq
\sum_{i=1}^4 \alpha_i=\sum_{i=1}^4 \beta_i=0. \label{ieq}
\eeq
We divide the particles into bins corresponding to spherical shells of thickness {{200}} pc each, and for the $j$th bin we calculate $\alpha_{ij}$ and $\beta_{ij}$.  Here $i$ labels the four terms in Eqs. (\ref{maina}) and (\ref{mainb}) and $j$ labels the bin, so that $\alpha_{ij}$ and $\beta_{ij}$ are just given by $\alpha_i$ and $\beta_i$ evaluated using the results of GALIC for the $j$th bin.  The parameters $\alpha_{ij}$ and $\beta_{ij}$ are proportional to the particle density $\nu$ and so are smaller at high radii, where the particle number falls.  Here the derivatives in Eqs.~(\ref{maina}) and (\ref{mainb}) are evaluated using the discrete difference between neighboring radial bins.

To obtain a measure of the deviation from equilibrium which is independent of such normalizations, we define the dimensionless index
\beq
X_j=\frac{\left(\sum_{i=1}^4 \alpha_{ij}\right)^2}{\sum_{i=1}^4 \alpha_{ij}^2}+\frac{\left(\sum_{i=1}^4 \beta_{ij}\right)^2}{\sum_{i=1}^4 \beta_{ij}^2}. \label{xeq}
\eeq
We will often omit the index $j$.  For a {\bbf{generic}} configuration, one would expect all of the $\alpha_{ij}$ and also all of the $\beta_{ij}$ to be uncorrelated.  In this case the cross terms in the numerator of $X$ would {\bbf{be of the same order as the squared terms}} and so $X$ would be the sum of two terms each {\bbf{of order unity}}.  Therefore one would expect $X$ {\bbf{itself to be of order unity.  In Subsec.~\ref{hernsez} we will see that this is indeed the case for a system far from equilibrium}}.  On the other hand, a lower value of $X$ implies that the terms on the right hand sides of Eqs.~(\ref{maina}) and (\ref{mainb}) nearly cancel each other.

If Sec.~\ref{jeansez} is correct, then $X$ should approach 0 as a system tends to equilibrium{\bbf{\ (where Eq.~(\ref{ieq}) holds)}} and so as the number of steps in the spherically symmetric simulations increases.  {\bbf{On the other hand, for an aspherical equilibrium system we have argued that the right hand sides of Eqs.~(\ref{maina}) and (\ref{mainb}) are arbitrary functions of position.  These right hand sides are integrated into the radial bins $\sum_i\alpha_{ij}$ and $\sum_i\beta_{ij}$, and so these will be arbitrary functions of the bin index $j$.  As a result, $X_j$ will generically be nonzero even for equilibrium aspherical configurations.  Therefore, as an aspherical system approaches equilibrium\ }}one expects $X$ to tend to a nonzero {\bbf{function}} corresponding to the systematic error caused by the unwarranted assumption of spherical symmetry.

\subsection{Results}

{\bbf{Each simulation produces a near equilibrium configuration of $10^5$ gravitationally interacting particles.  These particles can be stellar or dark matter.  To compare with observations, one needs to decide which particles are dark matter and which are stellar.  This must be done such that each population is separately near equilibrium.  We will define this segregation by simply imposing that each particle with specific energy less than $-1000\ ({\rm km/sec})^2$ is a stellar tracer.  As specific energy is a conserved quantity, such subpopulations of an equilibrium system will each be in equilibrium given the total gravitational potential.  With this tracer condition, each simulation produces about $1.8\times 10^4$ stars.  As will be explained in Sec.~\ref{tmtsez}, this corresponds roughly to the number of stars per classical dSph whose proper motions will be measured by upcoming extremely large telescopes.  The radial distribution of the tracers is also deep inside of the dark matter halo, as is expected for a typical dSph.}}

{\bbf{The tracer condition is of course not realistic, most obviously because the stars are far too massive.  However it is computationally cheap and we believe that this crude approximation does not affect our conclusions regarding the effectiveness of the parameter $X$ as a test of equilibrium.}}

\begin{figure} 
\begin{center}
\includegraphics[width=2.8in,height=1.6in]{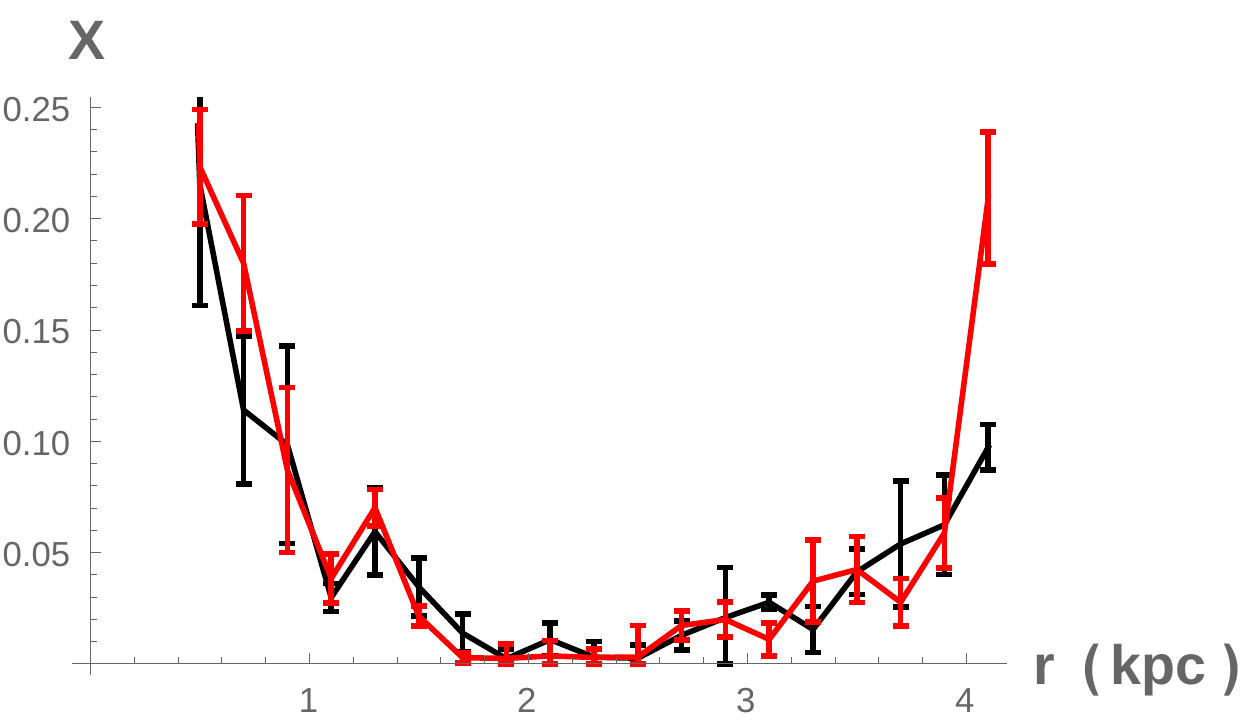}
\includegraphics[width=2.8in,height=1.6in]{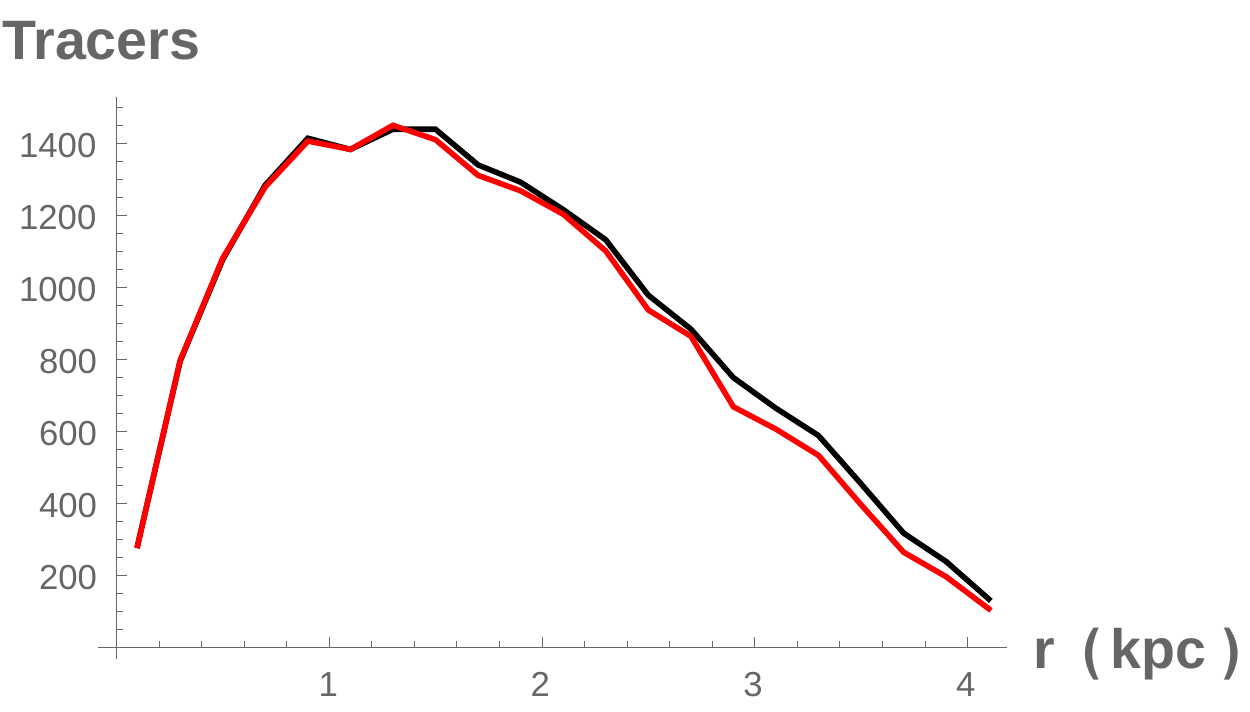}
\includegraphics[width=2.8in,height=1.6in]{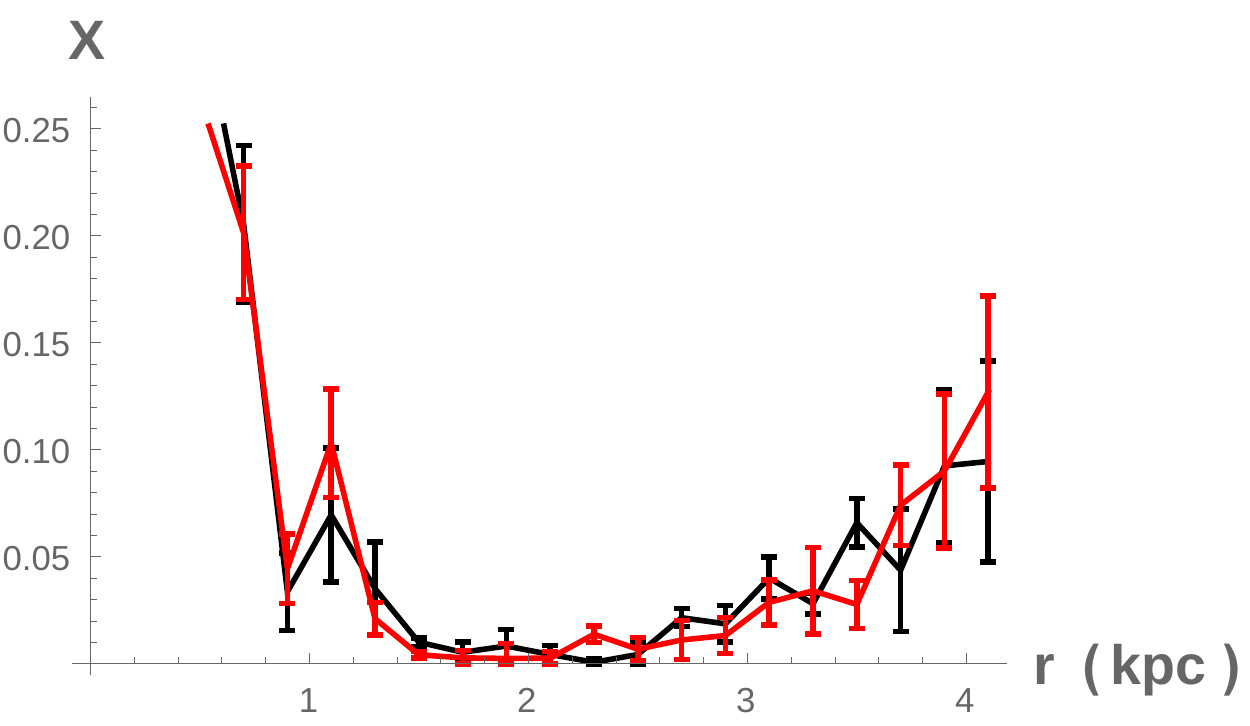}
\includegraphics[width=2.8in,height=1.6in]{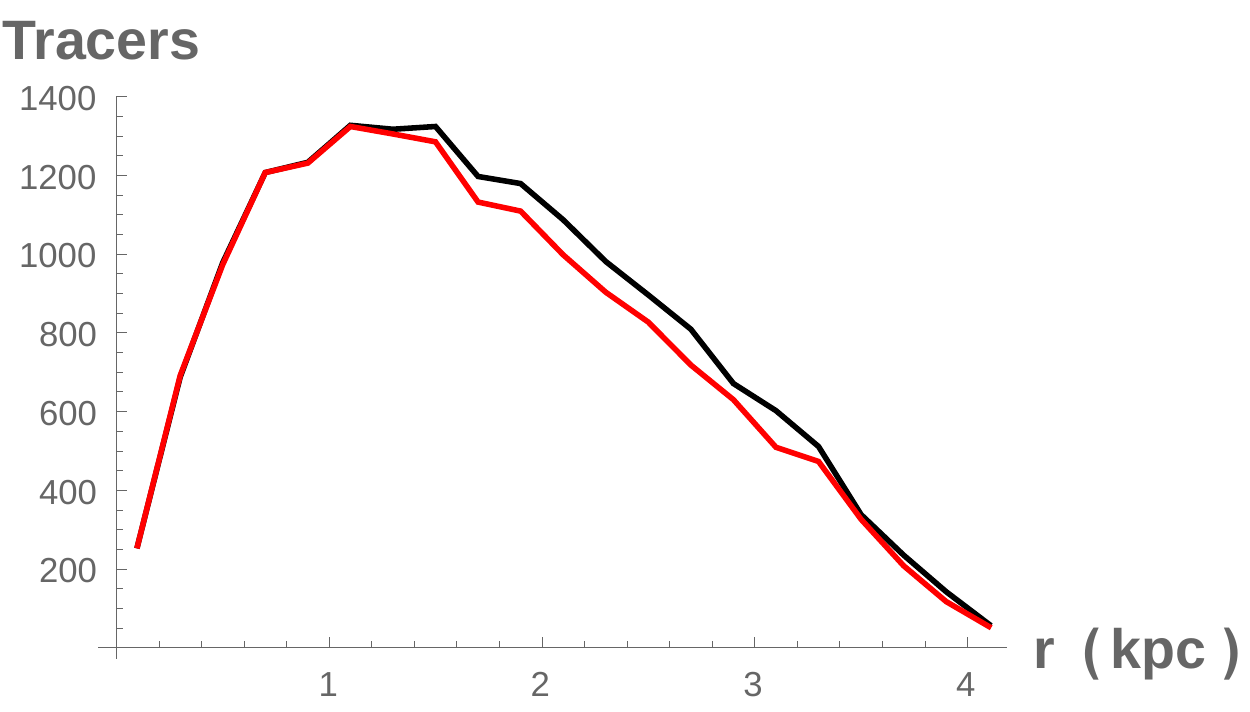}
\includegraphics[width=2.8in,height=1.6in]{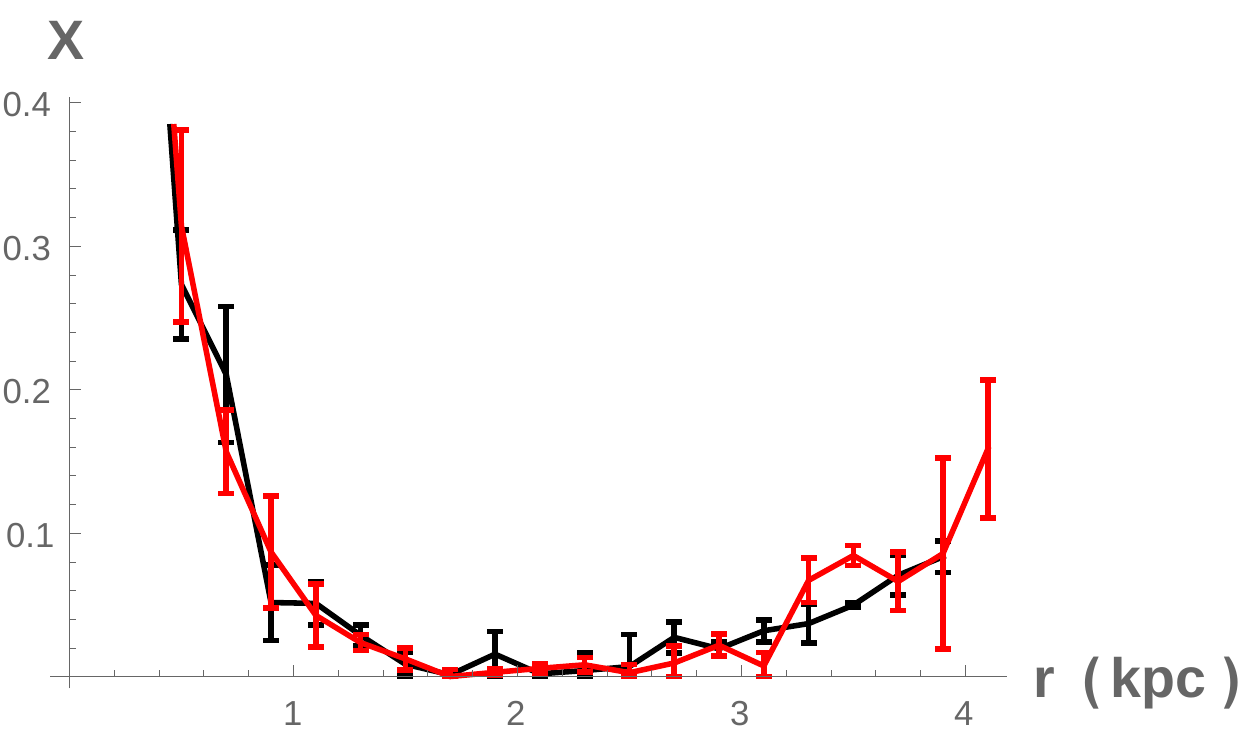}
\includegraphics[width=2.8in,height=1.6in]{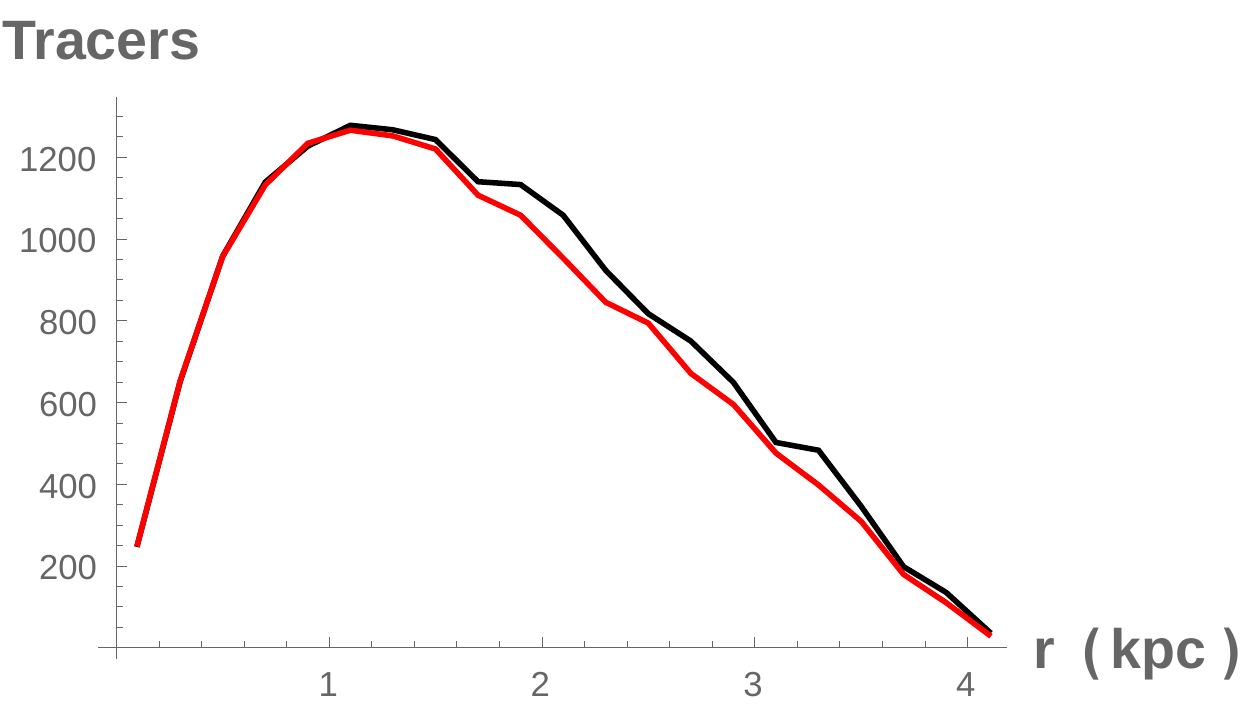}
\caption{The panels represent a spherically symmetric (top) and two prolate axisymmetric configurations, with axis ratios of $0.85$ (middle) and $0.8$ (bottom).  {\bbf{The two curves correspond to running GALIC with $0$ (black) and $500$ (red) steps/particle.  The right panels count particles with specific energies beneath $-1000\ ({\rm km/sec})^2$.  The left panels plot $X$ for each {{200}} pc radial bin.  The error bars reflect only statistical fluctuations.  A value $X\sim 1$ }}would correspond to a {\bbf{generic nonequilibrium}} configuration, the small values of $X$ imply that the tracers already begin near equilibrium, and stay near equilibrium. } 
\label{resfig}
\end{center}
\end{figure}

In Fig.~\ref{resfig} we plot our results for the value of $X$ in all {{200}} pc bins up to 4.5 kpc, {\bbf{beyond which the tracer condition cannot be satisfied}}.   In the inner 800 pc and also beyond 3 kpc, the number of particles per bins is small, leading to considerable shot noise which is compounded by the large bin size.  {\bbf{The large bin size is an important source of error because}} the derivatives are calculated using differences in neighbouring bins.  The top, middle and bottom panels represent a spherically symmetric, isotropic density distribution and two prolate axisymmetric distributions with eccentricities of $0.85$ and $0.8$.  {\bbf{The number of tracers per bin in each case is plotted in the panels on the right.  The lowest values of $X$ are achieved at radii of 1.5 to 3 kpc, where each bin contains a large number of particles and also the derivatives are relatively small and so well approximated by the differences between neighboring bins.  The error caused by finite binning is estimated in Subsec.~\ref{hernsez}.}}

{\bbf{The error bars represent only statistical fluctuations, not the finite binning uncertainties.  The statistical fluctuations are estimated by partitioning the tracers and calculating $X$ for each partition.  Then the uncertainty is estimated as the square root of the variance of the $X$ values of these subsamples divided by the square root of the number of partitions.  We have checked that this value is independent of the number of partitions so long as there are few enough partitions that the fluctuations are smaller than the mean value.  As the finite binning uncertainties are not included, we expect these errors to be grossly underestimated at small radii and also near 4 kpc.}}

{\bbf{The main result of this section is the fact that $X$ indeed is quite close to 0, as compared with the expected nonequilibrium value of order unity, when these errors are small.  In general it is of order $10^{-2}$ in this case of roughly $10^4$ tracers.  Interestingly, this remains true even in the strongly prolate configurations.  Thus the breakdown of spherical symmetry does not appear to significantly compromise the ability of $X$ identify equilibrium configurations.}}

{\bbf{There is little evolution in $X$ between the 0th and 500th steps of the relaxation, which indicates that this tracer population already begins quite close to equilibrium.  We will see in Subsec.~\ref{tuttipart} that this is not true for all choices of tracer.  This is because higher energy tracers, which were removed by the energy cut in this subsection, more often sample the nongaussian part of the Hernquist distribution, whereas GALIC configurations begin with purely Gaussian velocity distributions.  Thus the higher energy particles in GALIC begin further from equilibrium.  In a real galaxy one does not know {\it{a priori}} which subpopulation best approximates equilibrium, however one may use the methods of this paper to measure $X$ for each.  The subpopulation with the lowest value of $X$ may be the closest to equilibrium and so the most reliable for the construction of a mass model.}}

\begin{figure} 
\begin{center}
\includegraphics[width=2.8in,height=1.6in]{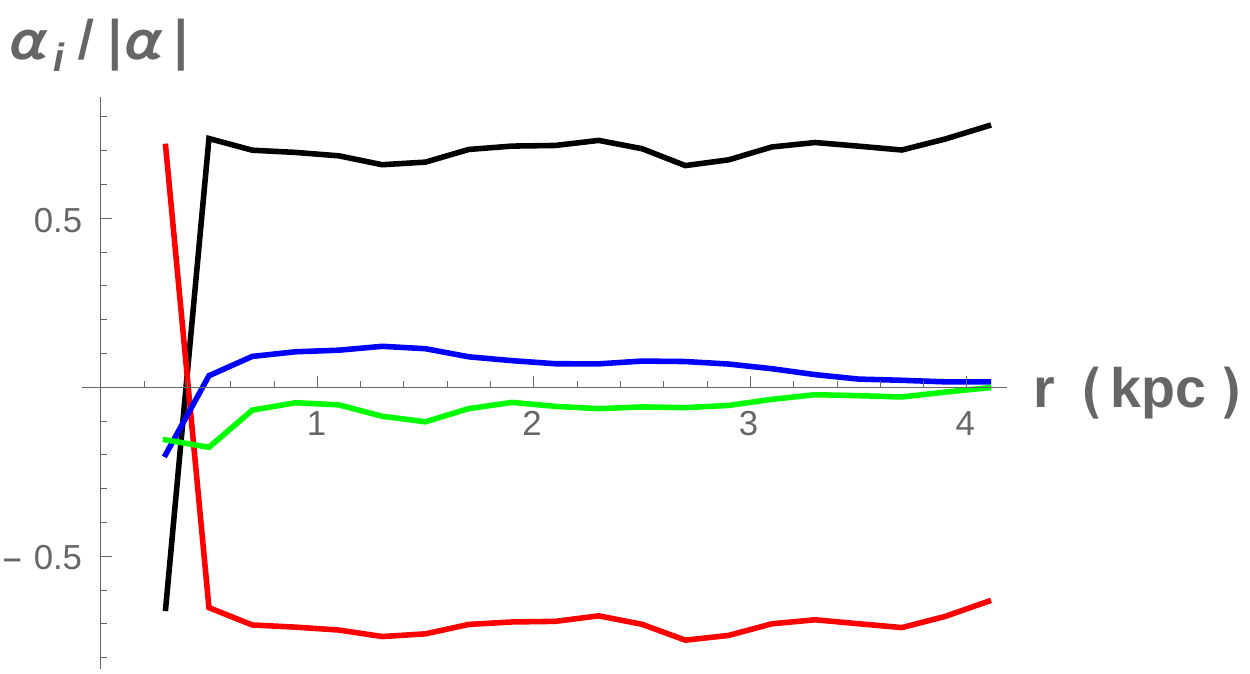}
\includegraphics[width=2.8in,height=1.6in]{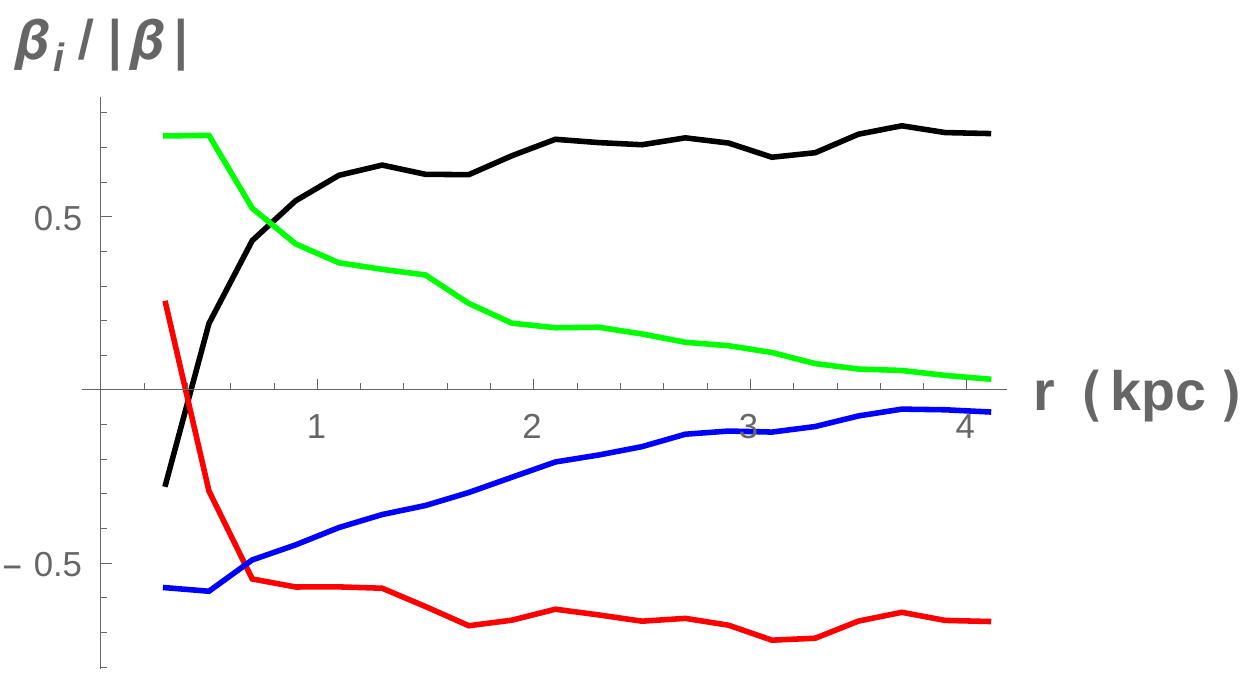}\\
\includegraphics[width=2.8in,height=1.6in]{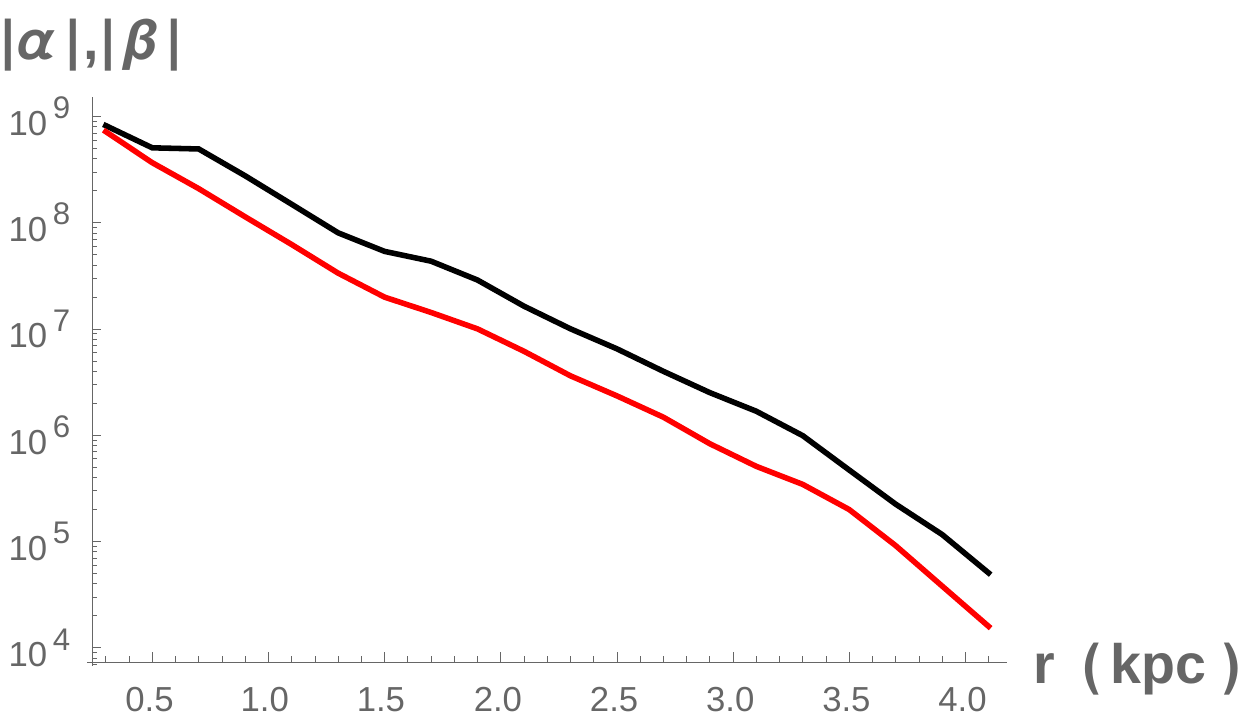}
\caption{The panels represent the individual contributions to $X$ in the case of the spherically symmetric configuration after {\bbf{500}} steps/particle.  The upper panels are the functions $\alpha_i/\sqrt{\sum_i\alpha_i^2}$ and $\beta_i/\sqrt{\sum_i\beta_i^2}$ respectively.  The black, red, blue and green curves represent the terms $i=1,\ 2,\ 3$ and $4$ respectively.   The lower panel plots the bin-dependent normalizations $\sqrt{\sum_i\alpha_i^2}$ in black and $\sqrt{\sum_i\beta_i^2}$ in red.}
\label{contribfig}
\end{center}
\end{figure}

The individual terms in $X$ are displayed in Fig.~\ref{contribfig} in the case of the spherically symmetric halo.  {\bbf{We can see that while individual terms increase or decrease their relative significance as a function of radius, nonetheless the equilibrium condition always conspires to make them nearly cancel.}}

\subsection{Aspherical Halos}


The formulae presented in this paper were derived assuming a spherically symmetric halo, and so one only expects $X$ to approach zero in this case.  The purpose of the simulation of the prolate halos is to determine whether the resulting increase in $X$ is sufficiently dramatic to render $X$ insensitive to the distance between the dark matter configuration and equilibrium.  {\bbf{On the contrary, in Fig.~\ref{resfig} one can observe that the eccentricity has only a modest effect on $X$, with the highly prolate bottom panel displaying an $X$ which is larger than that in the spherical top panel by only about one standard deviation in each bin.  As a result we are unable to quantify the effect of asphericity on the $X$ value of the tracer population at this time.}} In Subsec.~\ref{tuttipart} we will present an example in which the eccentricity has a larger effect on $X$.

What about realistic dark matter halos?  The GALIC code can only generate ellipsoidal halos.  However in $\Lambda$CDM simulations halos are triaxial \citep{frenkasse} with radius-dependent axis ratios \citep{hayashiasse}.  The triaxiality and radial-dependence may provide serious complications to certain quantitative tasks, such as determining the shape of the dark matter halo of a galaxy cluster from the shapes of X-ray isophotes \citep{hayashiasse}.  However, we expect that the correlation between $X$ and the equilibrium condition increases monotonically with the axis ratios.  {\bbf{In other words, the closer the axis ratios are to unity, the more accurately $X$ reflects the distance from equilibrium.}}  Therefore, for the purpose of estimating the impact of asphericity on $X$ in a certain radial bin, it is sufficient to consider a typical axis ratio that occurs in the line of sight integral corresponding to that bin.

The Aquarius simulations \citep{aquarius} of the evolution of pure dark matter halos were used to study subhalo shapes in \citet{ellip}.  The authors then used the semianalytical method of \citet{starkenburg} to determine which subhalos were likely to host local group satellite galaxies.  The authors found that these had {\bbbf{minor to major}} axis ratios ranging from $0.4$ at about 1 kpc to $0.8$ at the radius which the authors called $r_{95}$, equal to about $2/3$ of the Virial radius.  In our case, at the smallest radii, $X$ is dominated by binning effects and in particular the imprecision in our discrete derivative.  At the radius $r_{95}$ {\bbbf{the authors}} also estimated the intermediate {\bbbf{length principal axis}} to major axis ratio and found about $0.9$, suggesting that axisymmetry is a reasonable approximation.


\subsection{Time scales and dimensional analysis}

As these systems are not in equilibrium, the velocity moments evolve with time.  In this subsection we will use a dimensional analysis of the Jeans Eqs.~(\ref{maina}) and (\ref{mainb}) to relate the quantity $X$ to the time scale $t$ over which the velocity moments evolve.

Let us make the crude approximation that each term in Eq.~(\ref{xeq}) is equal to $X/2$.  Now each term on the right hand sides of Eqs.~(\ref{maina}) and (\ref{mainb}) is of order $v^4/r$ where $v$ is a characteristic stellar velocity, say 5 km/s for a UFD or 10 km/s for a classical dSph, and $r$ is the radial coordinate, which is of order 100-300 pc for known dSphs.  Thus the denominator of each term in Eq.~(\ref{xeq}) consists of four terms, each of which is about $v^8/r^2$.  For simplicity we add these four terms in quadrature in what follows, as if they were of comparable magnitudes.  On the other hand, the total right hand side is equal to the left hand side, which is of order $v^3/t$ where $t$ is the characteristic timescale over which the nonequilibrium system evolves.  For example, if the system were in equilibrium then $t$ would be infinite.

One can then estimate the order of each term in $X$ as
\beq
\frac{X}{2}\sim\frac{v^6/t^2}{\sqrt{4} v^8/r^2}=\frac{r^2}{2v^2 t^2}\sim \frac{1}{32}\frac{p^2}{t^2}
\eeq
where $p$ is the period of a typical star, which we have roughly estimated to be $4r/v$.  We can now interpret $X$ as giving the timescale over which the nonequilibrium system significantly evolves
\beq
t\sim \frac{p}{4\sqrt{X}} 
\eeq
which for $X\sim 10^{-3}$ would be of order one Gyr.  This means that over a period of $\epsilon$ Gyr, if $\epsilon$ is sufficiently small then one would expect a fractional change in the {\bbf{odd}} velocity moments of order $\epsilon$.  {\bbf{As a result, if $X$ is far from 0 but the first and third velocity moments are still quite small, this suggests that the system {\bbbf{had been in equilibrium until it}} was disrupted roughly within the past Gyr.}}

{\bbf{It is tempting to attempt to compare the disruption time estimated here with the relaxation time of a system relaxing in GALIC.  However we know of no direct relation between the relaxation steps of the GALIC code and the real time evolution of a relaxing system, indeed it is unlikely that any such relation exists as the relaxation rate in GALIC is optimized to run quickly, rather than matched to the dynamical relaxation rate.}}


\subsection{Application to All Particles} \label{tuttipart}

{\bbbf{In the examples thus far we have considered small deviations from equilibrium.  In these cases we have observed that}} the value of $X$ is independent of the number of steps, and so appears to be independent of the distance from equilibrium.  {\bbf{In Subsec.~\ref{hernsez} we will see that this is not true for larger deviations from equilibrium.}} In this subsection, to show that $X$ {\bbbf{can}} indeed depend on the distance from equilibrium {\bbbf{even for small perturbations from equilibrium}}, we will provide an example in which $X$ does evolve with the number of steps.  This example is the same as that presented in the previous section, except that instead of restricting our attention to tracers, we consider the full $10^5$ particles.  As these higher energy particles have a much greater spatial extent, we will use 4 kpc radial bins instead of {{200}} pc bins.

\begin{figure} 
\begin{center}
\includegraphics[width=2.8in,height=1.6in]{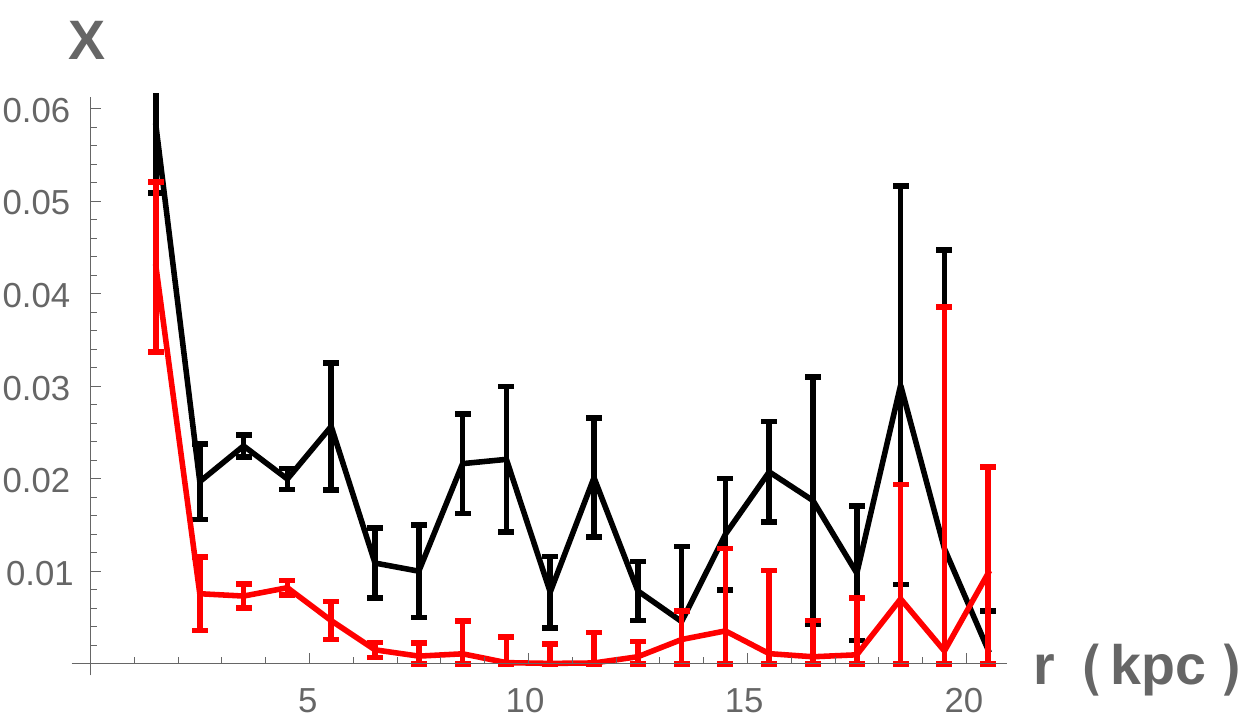}
\includegraphics[width=2.8in,height=1.6in]{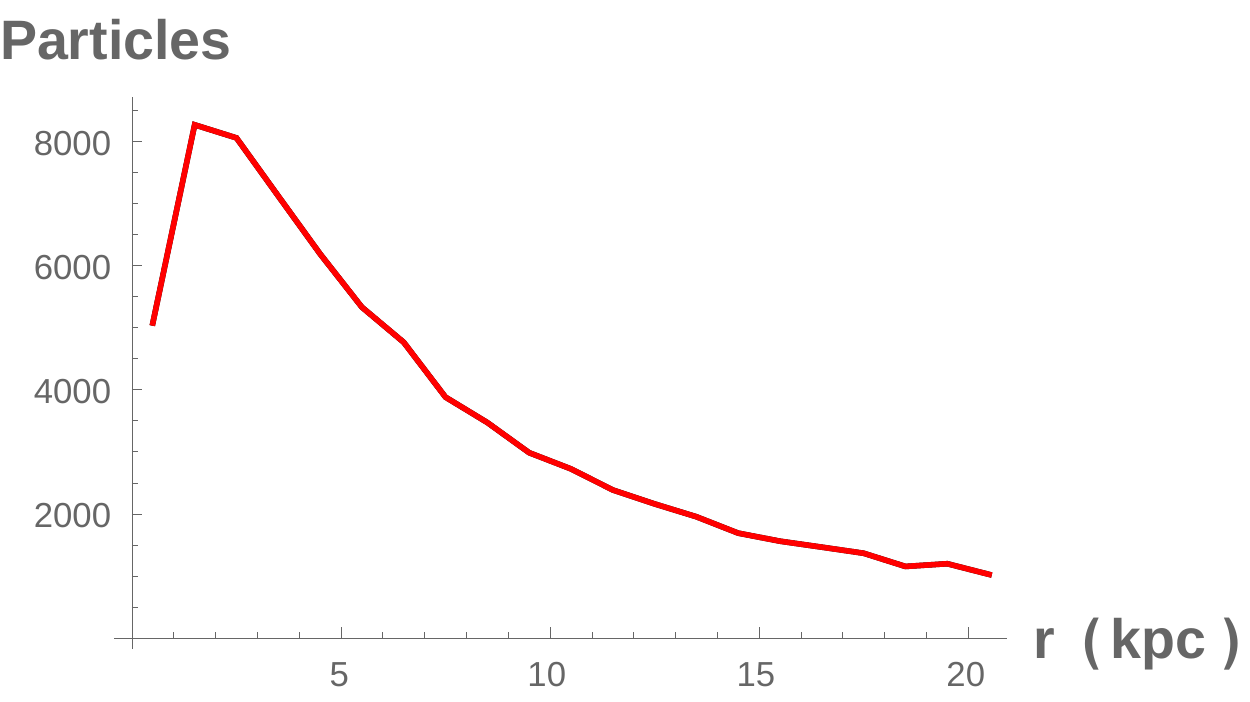}
\includegraphics[width=2.8in,height=1.6in]{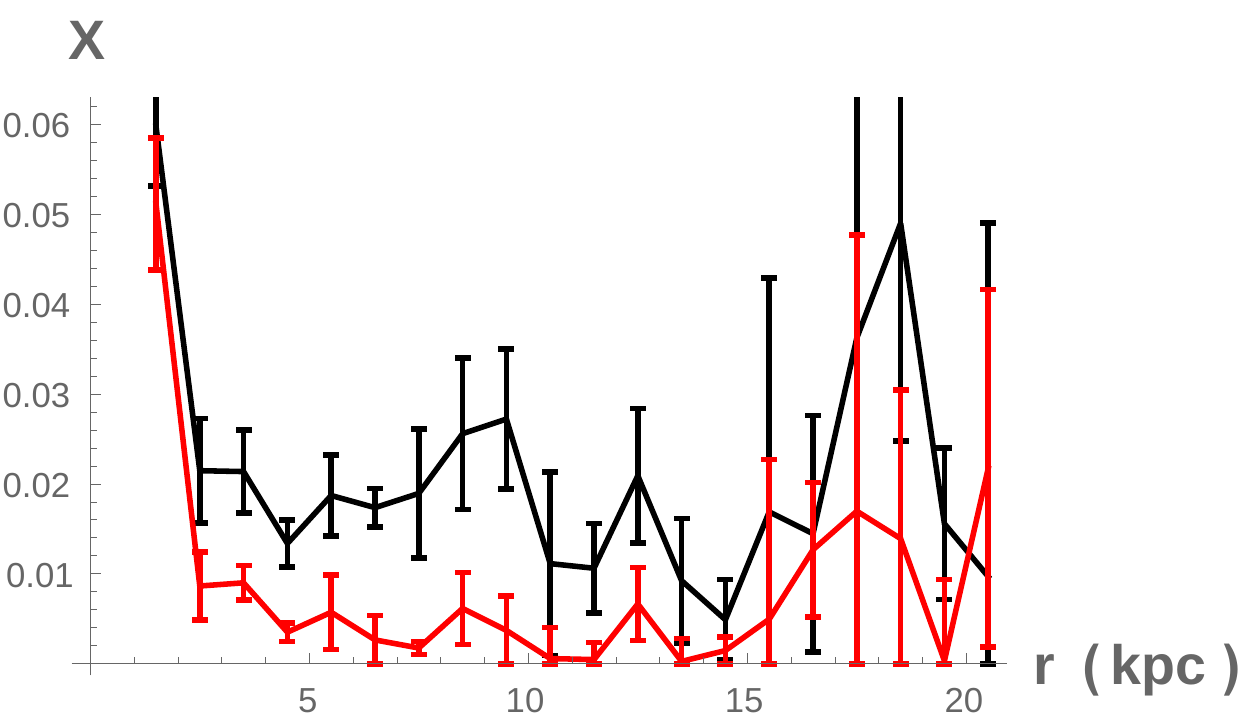}
\includegraphics[width=2.8in,height=1.6in]{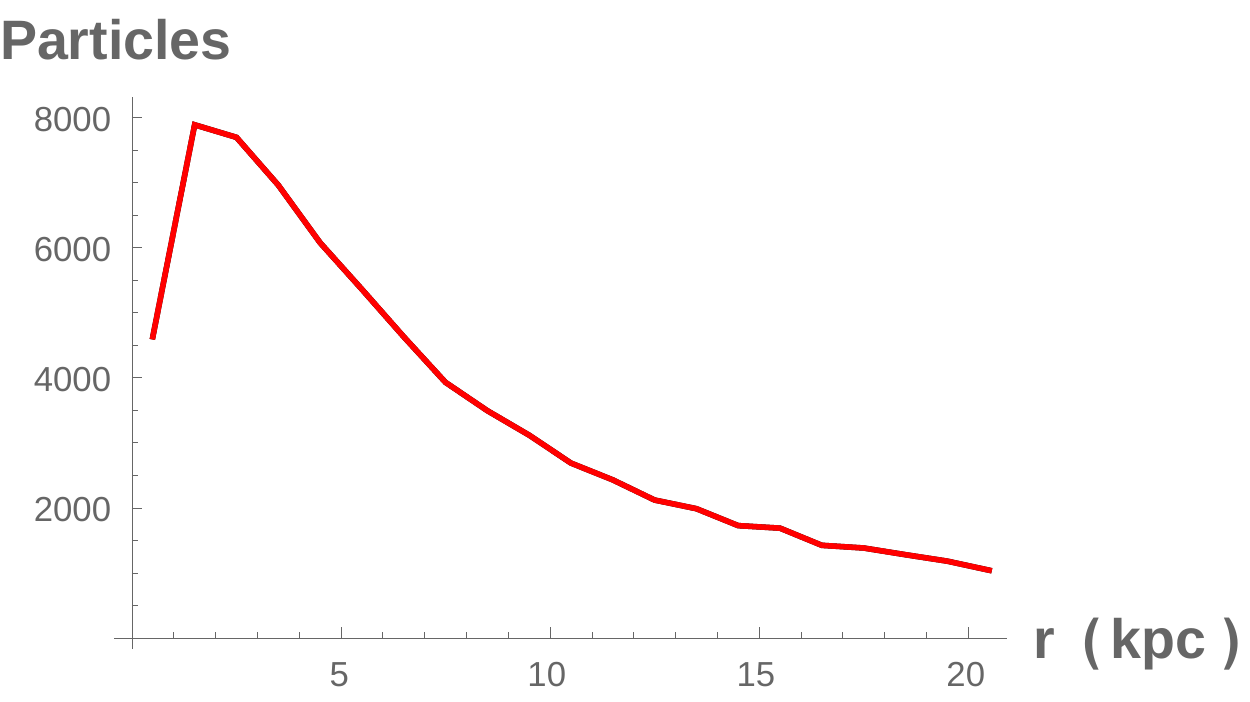}
\includegraphics[width=2.8in,height=1.6in]{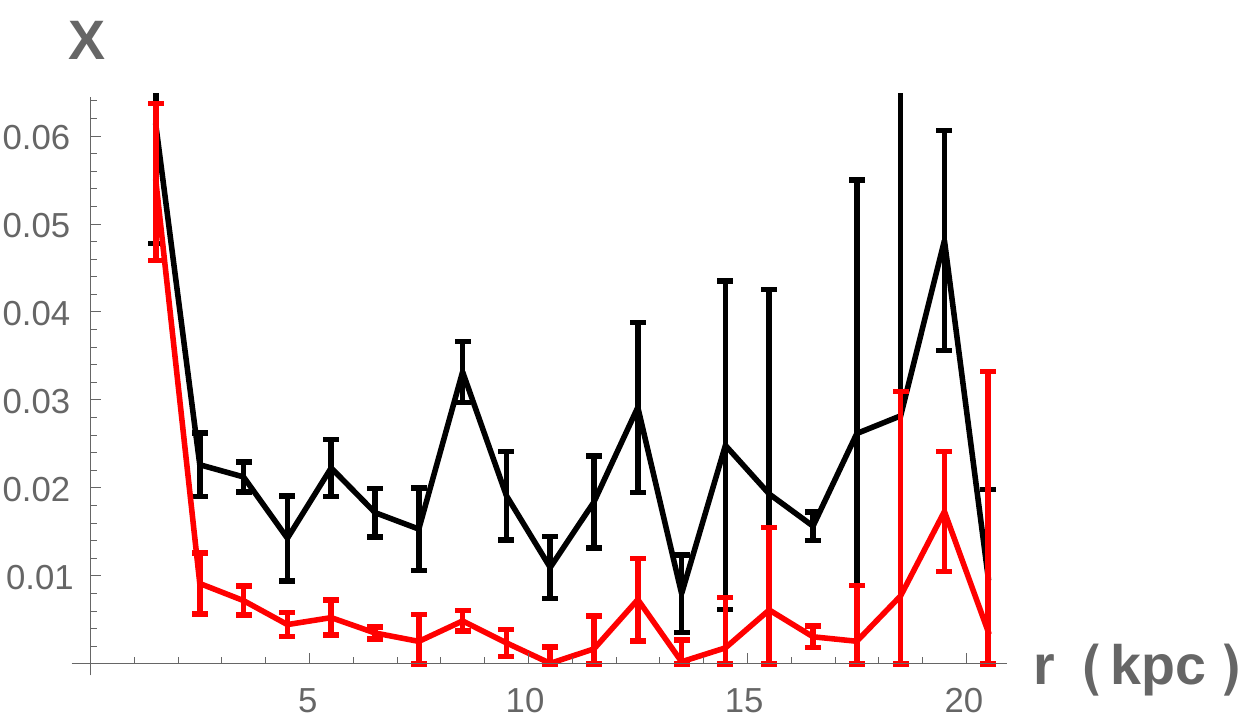}
\includegraphics[width=2.8in,height=1.6in]{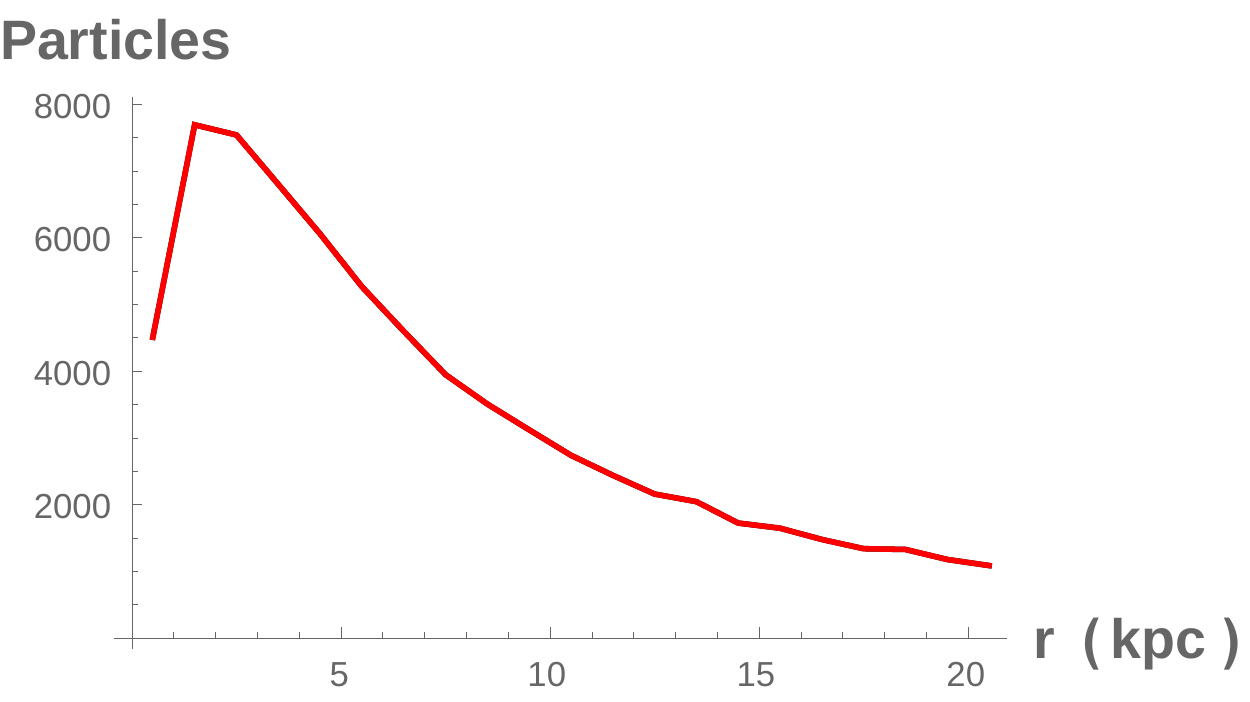}
\caption{As in Fig.~\ref{resfig} except that all $10^5$ particles are considered, not only the tracers.  Also the bin size has been increased to 4 kpc.  The number of particles per bin is independent of the number of the steps as GALIC uses a made-to-measure algorithm to generate the distributions, and so the spatial coordinates of the particles are fixed.}
\label{notracefig}
\end{center}
\end{figure}

Our results are shown in Fig.~\ref{notracefig}.  Here one can clearly observe that $X$ decreases as the number of steps increases, and so as the system relaxes towards equilibrium.  {\bbf{In the spherically symmetric case, $X$ reduces from about $2\times 10^{-2}$ to about $10^{-3}$ at intermediate radii, where the errors are smallest.}} This improvement is less evident in the more prolate cases, suggesting that $X$ saturates at some nonzero value, {\bbf{apparently between $2\times 10^{-3}$ and $5\times 10^{-3}$ at intermediate radii}}.  Such a saturation to some extent can be attributed to the systematic error induced by falsely assuming spherical symmetry, but it may also result from the fact that the GALIC code relaxes prolate configurations more slowly than spherically symmetric configurations{\bbf{, although the later possibility seems unlikely as we have observed little GALIC evolution between steps 100 and 500}}.  

\subsection{Tracers Drawn From Known Distribution Functions} \label{hernsez}

\begin{figure} 
\begin{center}
\includegraphics[width=4.2in,height=2.4in]{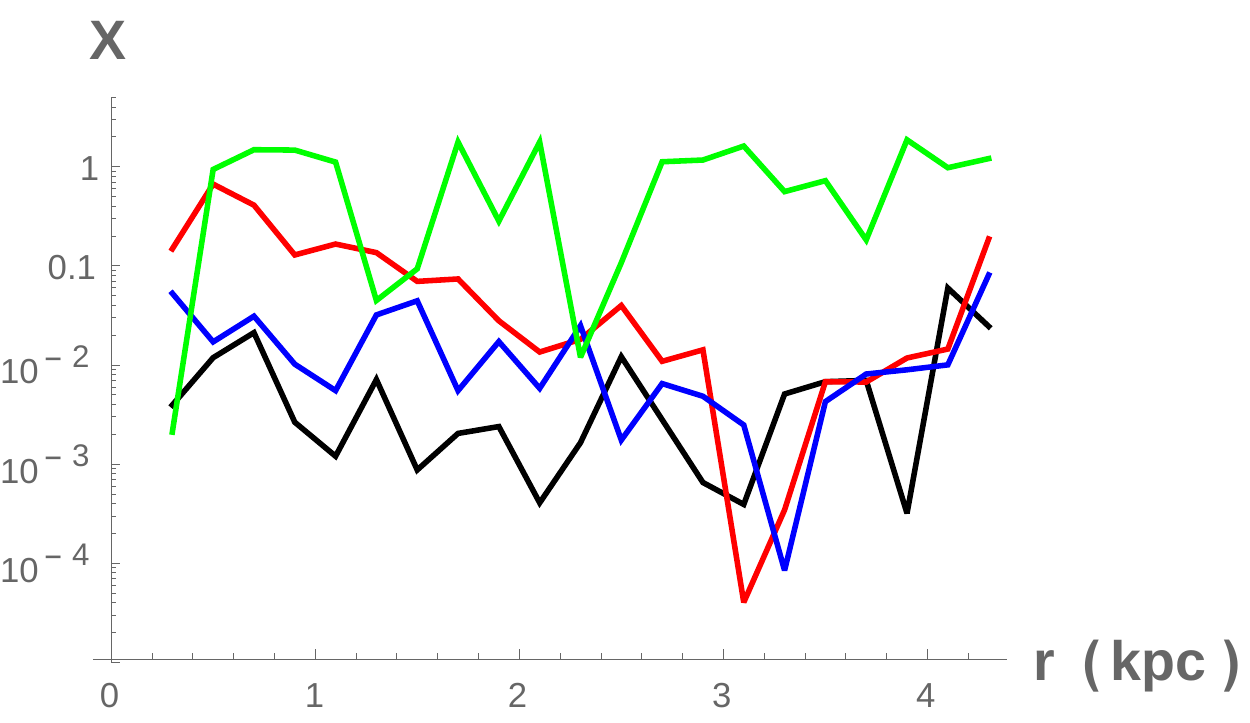}
\caption{This plot was generated by directly sampling {\bbf{$10^5$ particles from}} predefined spherically symmetric distribution functions, {\it{not}} using an $N$-body simulation.  In each case, the tracers are particles with specific energy less than $-1000\ ({\rm{km/sec}})^2$.  The four curves correspond to an isotropic Hernquist distribution (black), the same distribution with one component of the velocity multiplied by $1+r/(2a)$  (blue) and $1+2r/a$ (red) {\bbf{and also with all velocities multiplied by $1+2{\rm{sin}}\left(20r/a\right)$ (green)}}.  {\bbf{The first curve samples an exact, spherical equilibrium distribution.  Therefore the nonzero value of $X$ is entirely due to statistical fluctuations and finite binning.}}  The latter curves are further from equilibrium, and correspondingly $X$ is larger.}
\label{hernfig}
\end{center}
\end{figure}

{\bbf{In this subsection we do not use the $N$-body code.  Instead our tracers sample phase space distribution functions which we have fixed by hand.  This will allow several tests which would not be feasible with the full code.}}

We apply the same analysis as above to tracer populations drawn from {\bbf{four}} well-defined phase space distribution functions.  The first is a spherical, isotropic Hernquist profile with the same parameters as that used in the GALIC simulations.   {\bbf{This is our only example of a true equilibrium system.}}  The second and third are the same as the first but with one of component of the velocity at radial coordinate $r$ multiplied by $1+r/(2a)$ and $1+2a/r$ respectively, with $a$ equal to 3.45 kpc.  {\bbf{In the last, each velocity component is multiplied by $1+2{\rm{sin}}\left(20r/a\right)$.  This is our only example of a system which is far from equilibrium}}.  In each case $10^5$ particles were used with the same tracer condition, $E<-1000\ ({\rm{km/sec}})^2$, as in the simulated samples.

The results are shown in Fig.~\ref{hernfig}, where again the bin size is {{200}} pc.  The first profile is an equilibrium configuration.  The second configuration is not in equilibrium while the third is farther from equilibrium.  {\bbf{The last is very far from equilibrium and so as expected $X$ is of order unity.}} Thus $X$ in the first case is lower than in the GALIC configurations, which are close to equilibrium but still not in equilibrium.  One does not expect $X$ to be zero in any case due to the finite sample size and the corresponding finite binning.   As expected, the values of $X$ are larger in the nonequilibrium second configuration and even larger in the third, which is yet further from equilibrium.  

{\bbf{In practice the establishment of equilibrium requires a robust determination of the uncertainties entering $X$.  The first curve in Fig.~\ref{hernfig} shows the contribution to these errors arising from statistical fluctuations together with finite binning, but does not include asphericity.}}

\begin{figure} 
\begin{center}
\includegraphics[width=4.2in,height=2.4in]{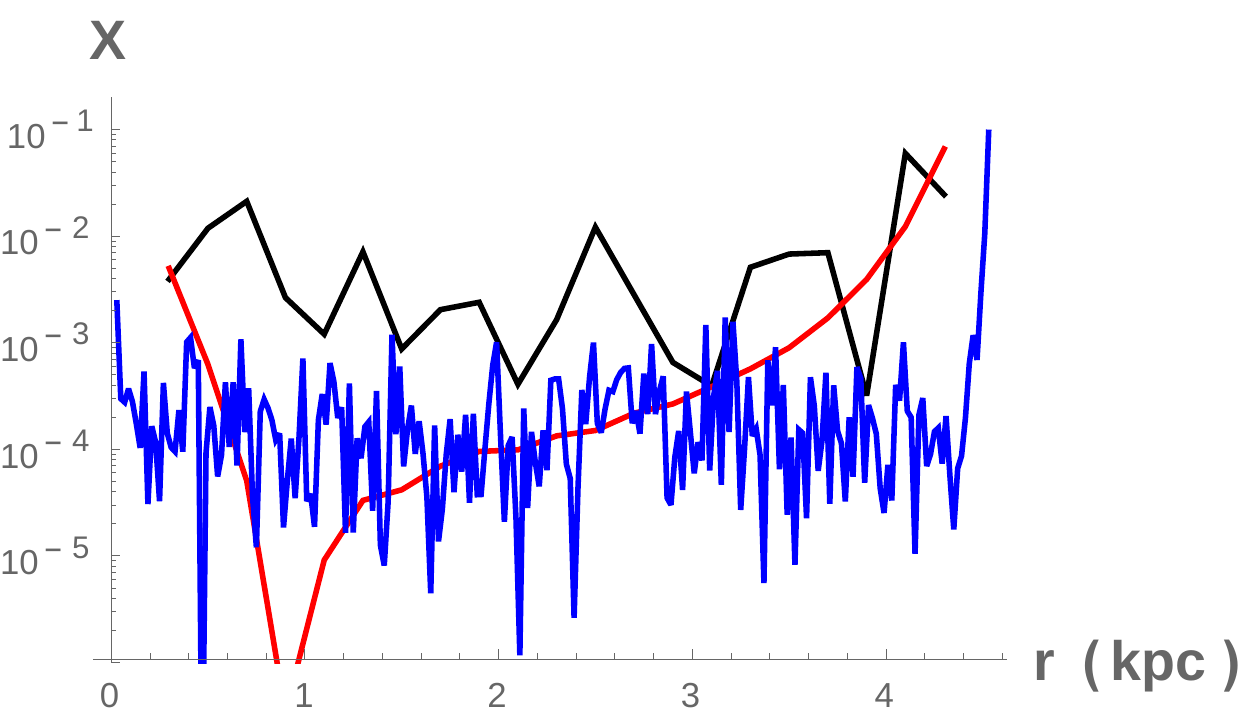}
\caption{{\bbf{As in Fig.~\ref{hernfig} but only the Hernquist distribution is sampled.  The black curve uses tracers from a $10^5$ particle sample with 200 pc bins, the red and blue curves draw tracers from a $10^9$ particle sample with 200 pc and 20 pc bins respectively.  The red curve provides an estimate of the error due to finite binning alone.}}}
\label{hernbillionfig}
\end{center}
\end{figure}

{\bbf{As these configurations are simply drawn from known phase space distributions, and not generated by $N$-body simulations, it is not computationally expensive to create configurations with as many as $10^9$ particles.  In Fig.~\ref{hernbillionfig} we compare the $10^5$ particle Hernquist configuration with two $10^9$ particle configurations, with bin sizes of 200 pc and 20 pc.  In each case, the same tracer condition is applied which selects 18\% of the particles.  }}

{\bbf{The $10^9$ particle calculations, as expected, have much lower values of $X$.  This of course is expected due to the small statistical fluctuations.  The 200 pc bin case in fact has negligible statistical fluctuations and so it reflects the errors that are expected from binning alone, not only in the present case but throughout the paper where 200 pc bins are used.   {\bbbf{In fact, due to the very high number of tracers per bin, this 200 pc bin curve is the only curve in the paper whose statistical fluctuations are so small that they are not apparent on the plot.}} As expected, the binning error becomes greatest near the origin and also near the tracer edge at $r=4.5$\ kpc, where the bin to bin fractional changes are large and so the finite bin differences are a poor approximation for the derivatives.}}

{\bbf{On the other hand, the 20 pc bin case is clearly dominated by statistical fluctuations.   While all $X$ plots in the paper in 200 pc bin cases turn upwards beneath 1 kpc, this 20 pc bin plot turns upwards only beneath 100 pc.  This confirms the fact that the high value of $X$ beneath 1 kpc in the other plots is due to a combination of finite binning and shot noise.  Here, with both of these contributions well below $10^{-3}$, one sees that the intrinsic value of $X$ indeed is below $10^{-3}$.  This is consistent with our central claim that in an equilibrium, spherical configuration, in the limit of a large number of tracers and perfect measurements, $X$ tends to zero.}}

\subsection{Projection} \label{projsub}

So far we have assumed that the full 3-dimensional positions and velocities are available to the observer.  In practice only the projection along the line of sight is available.  While the Thirty Meter Telescope will be able to measure the full 3-dimensional velocities for as many as $10^5$ stars in the brightest dSphs, the positions will nonetheless be projected onto the celestial sphere.

We have also considered the projections of the GALIC configurations described above, with the same specific energy condition at $-1000({\rm{km/sec}})^2$.  GALIC generates configurations with a rotational symmetry about the $z$-axis.  For concreteness, we have considered observations with the line of sight along the $x$-axis.  We then used the results of Sec.~\ref{projsez} to deproject these projected observables, yielding 3-dimensional observables which we used to calculate $X$ as above.  In the deprojection one needs to calculate derivatives of the projected moments.  These derivatives{\bbf{, like all derivatives of moments in this note}}, are calculated by averaging the discrete difference from neighboring radial bins with the discrete difference from next-to-neighbor radial bins.  An improved scheme for evaluating the derivatives would reduce the resulting fluctuations and as a result reduce $X$.  {\bbf{To improve the precision of the derivatives is quite easy, one need only average information from $n$th neighbor bins where $n>2$.  However this improved precision will lead to reduced accuracy if the functions in question are not sufficiently linear on the distance scale equal to $n$ times the bin size.  Thus a higher precision comes at a cost of less sensitivity to the fine structure of the mass profile.}}

\begin{figure} 
\begin{center}
\includegraphics[width=2.8in,height=1.6in]{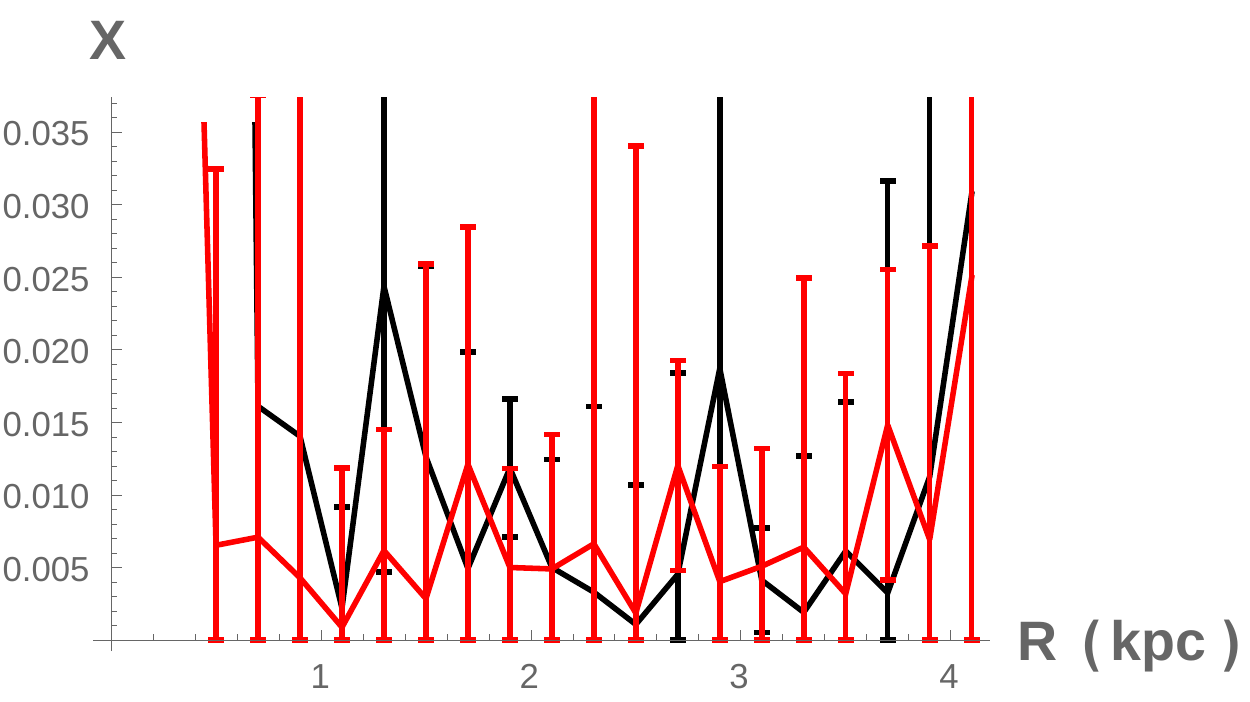}
\includegraphics[width=2.8in,height=1.6in]{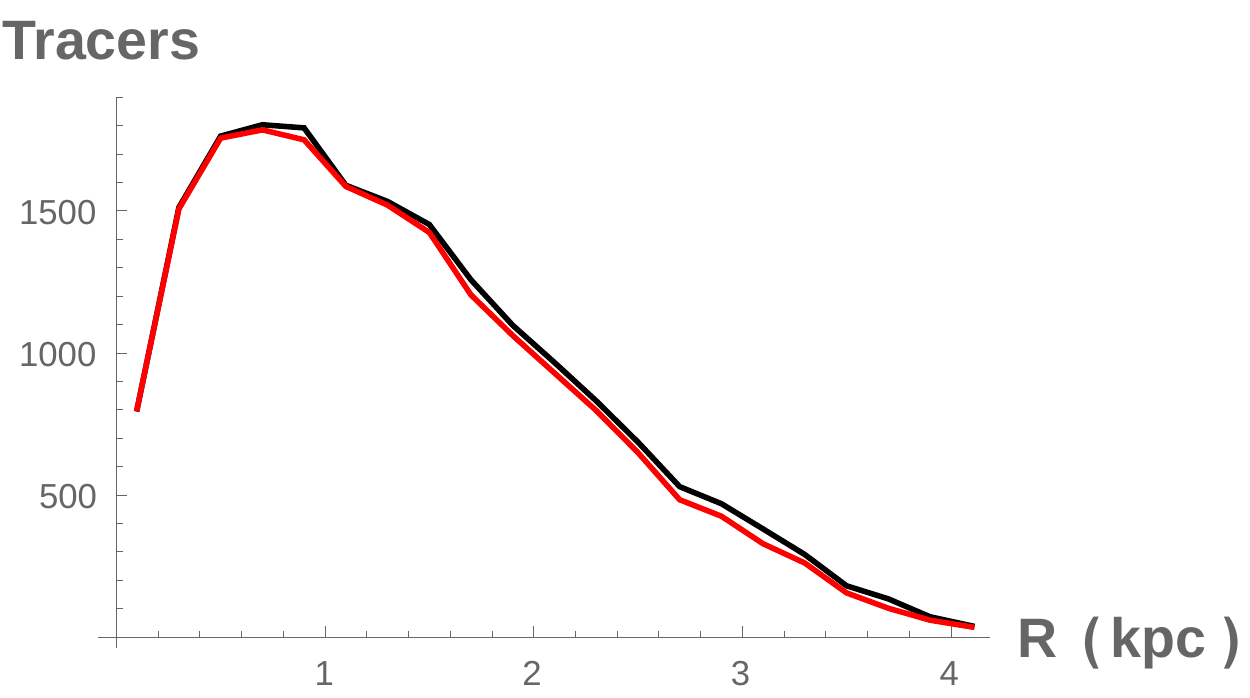}
\includegraphics[width=2.8in,height=1.6in]{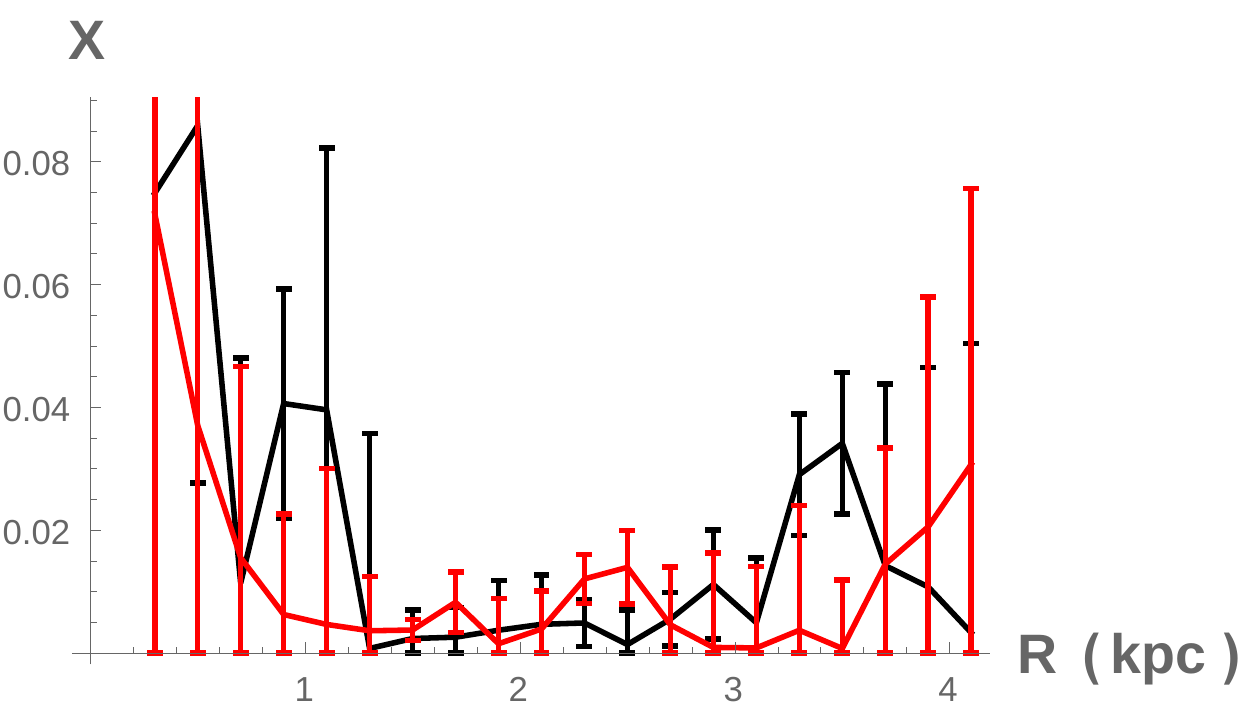}
\includegraphics[width=2.8in,height=1.6in]{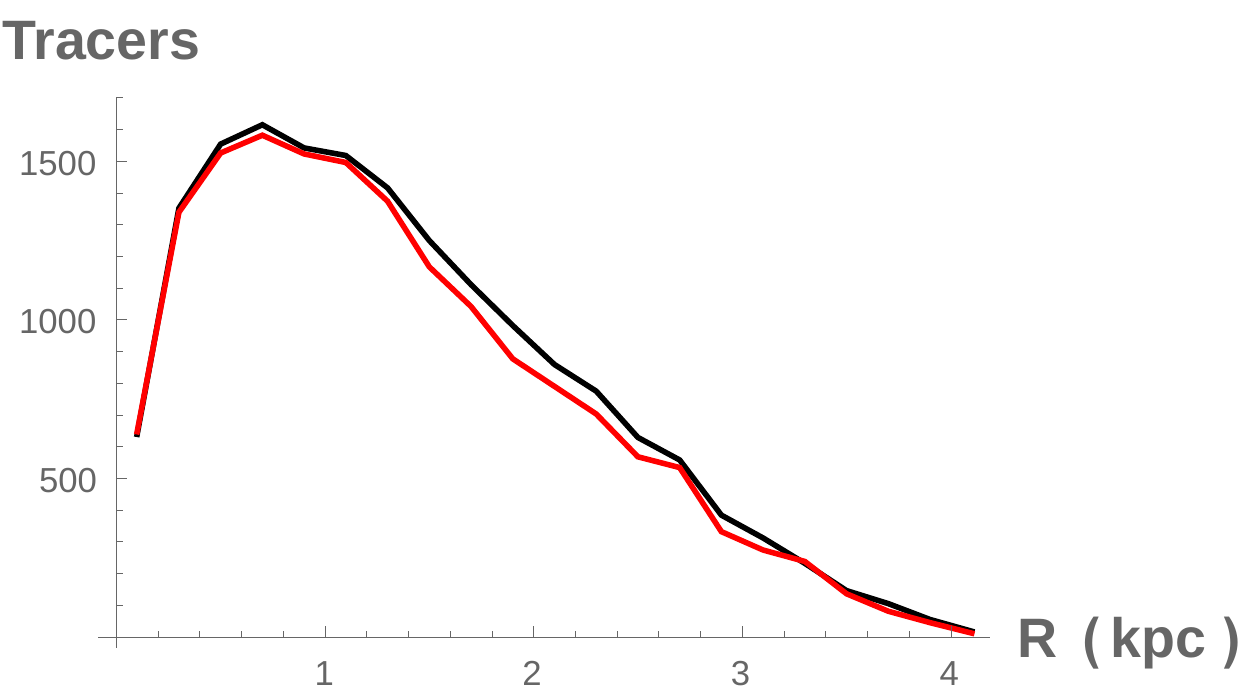}
\includegraphics[width=2.8in,height=1.6in]{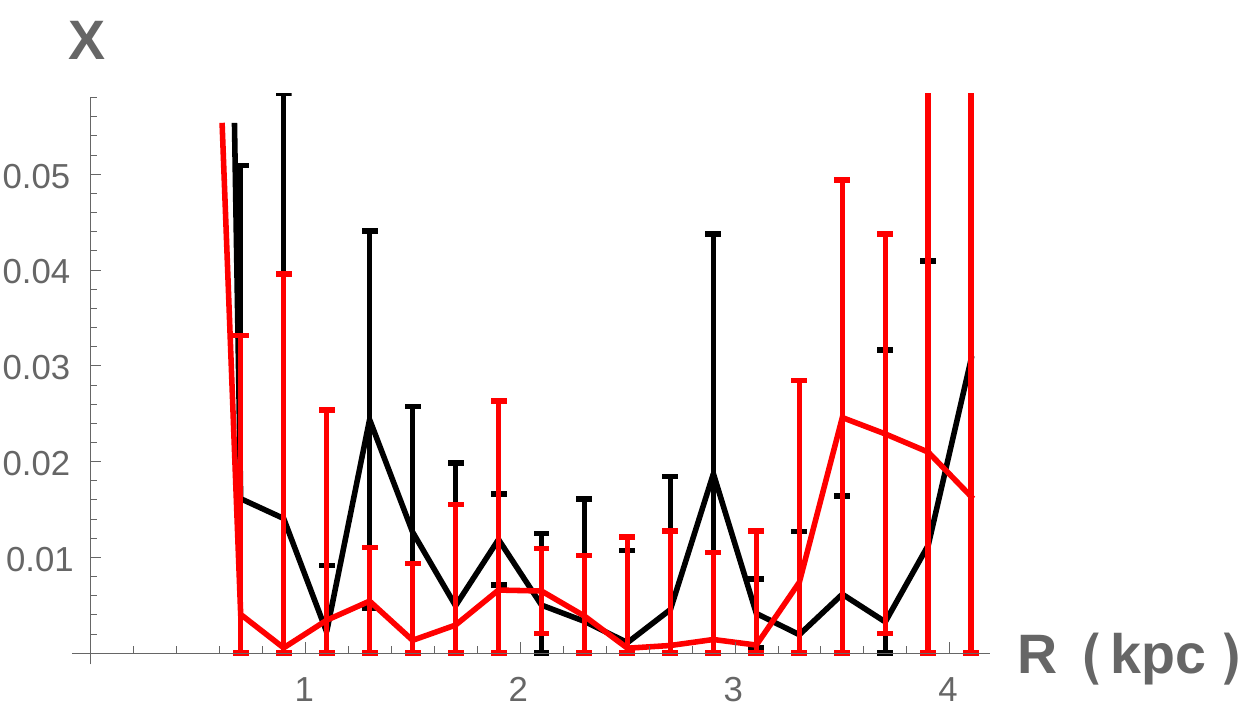}
\includegraphics[width=2.8in,height=1.6in]{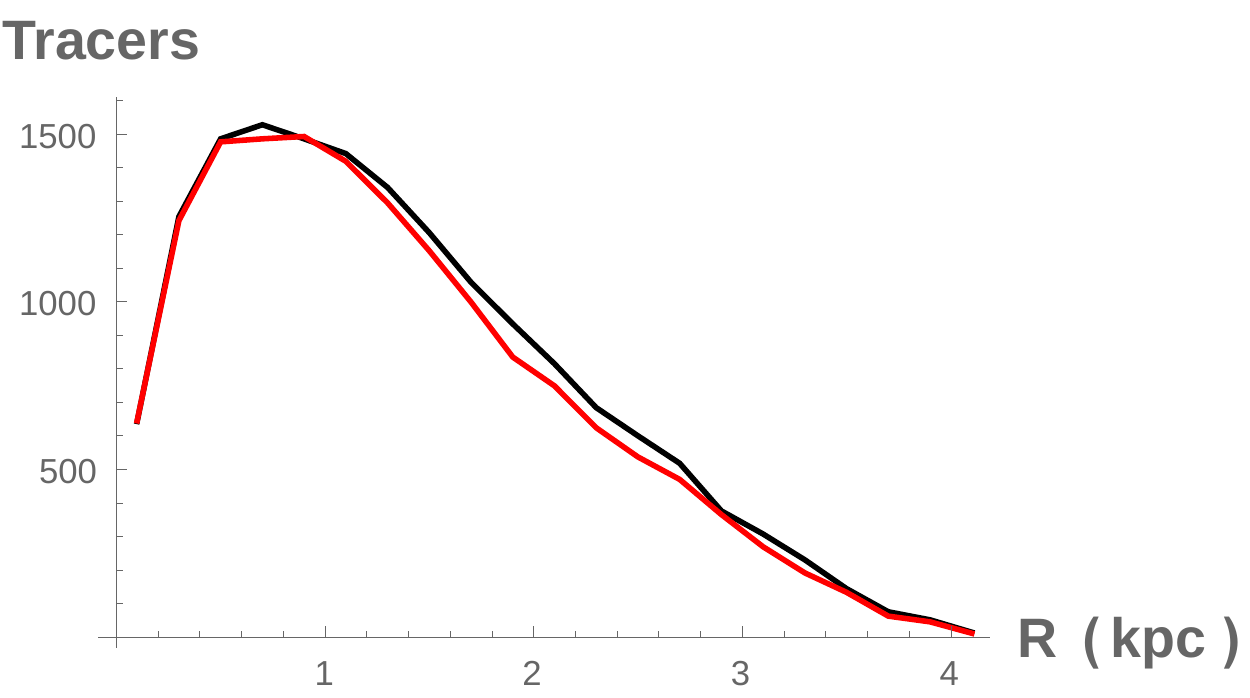}
\caption{The same three GALIC configurations of tracers as are considered as in Fig.~\ref{resfig}.  Only the line of sight projected information is used, where the line of sight is perpendicular to the symmetry axis of the halo.  These projected moments are deprojected using the inverse Abel transforms given in Sec.~\ref{projsez}.  Again each bin is 200 pc.}
\label{projfig}
\end{center}
\end{figure}
\begin{figure} 
\begin{center}
\includegraphics[width=2.8in,height=1.6in]{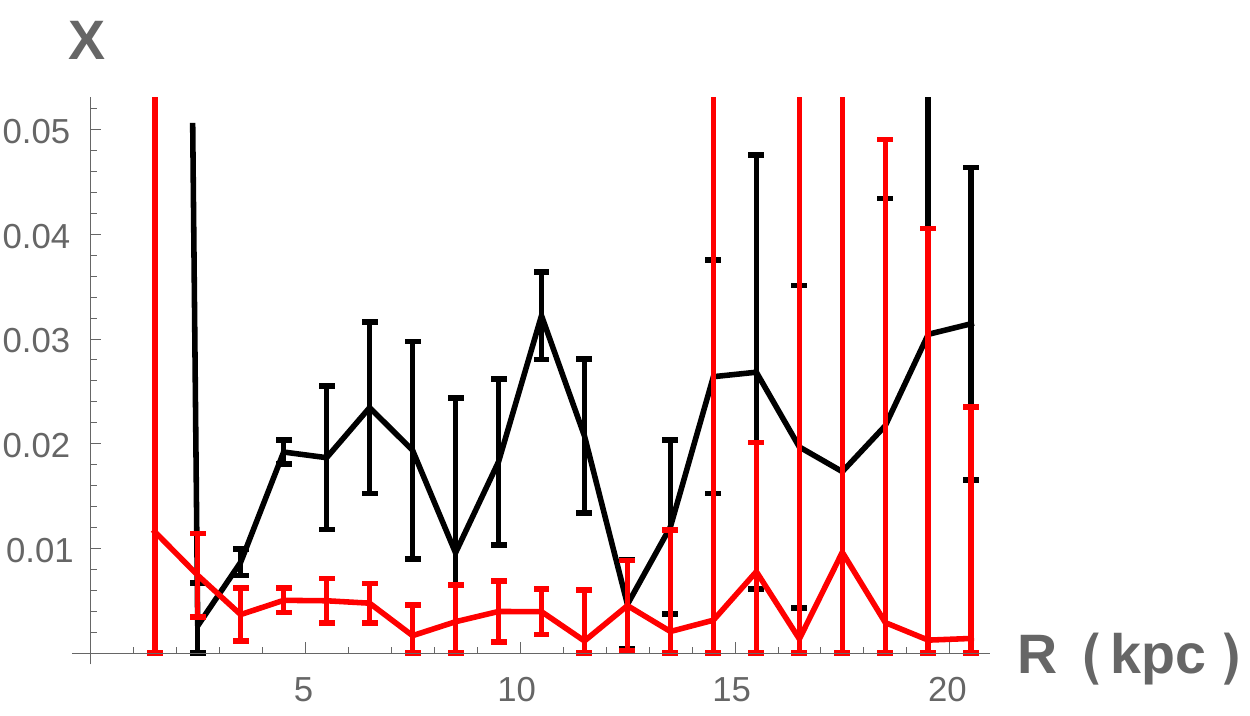}
\includegraphics[width=2.8in,height=1.6in]{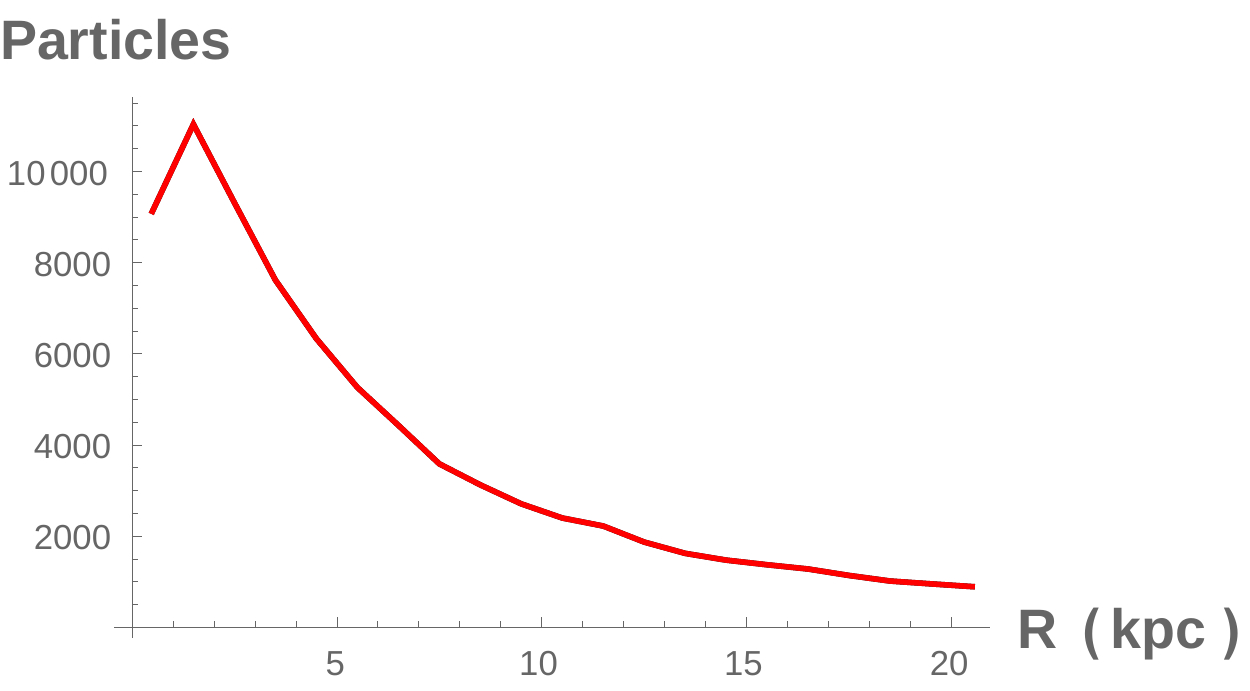}
\includegraphics[width=2.8in,height=1.6in]{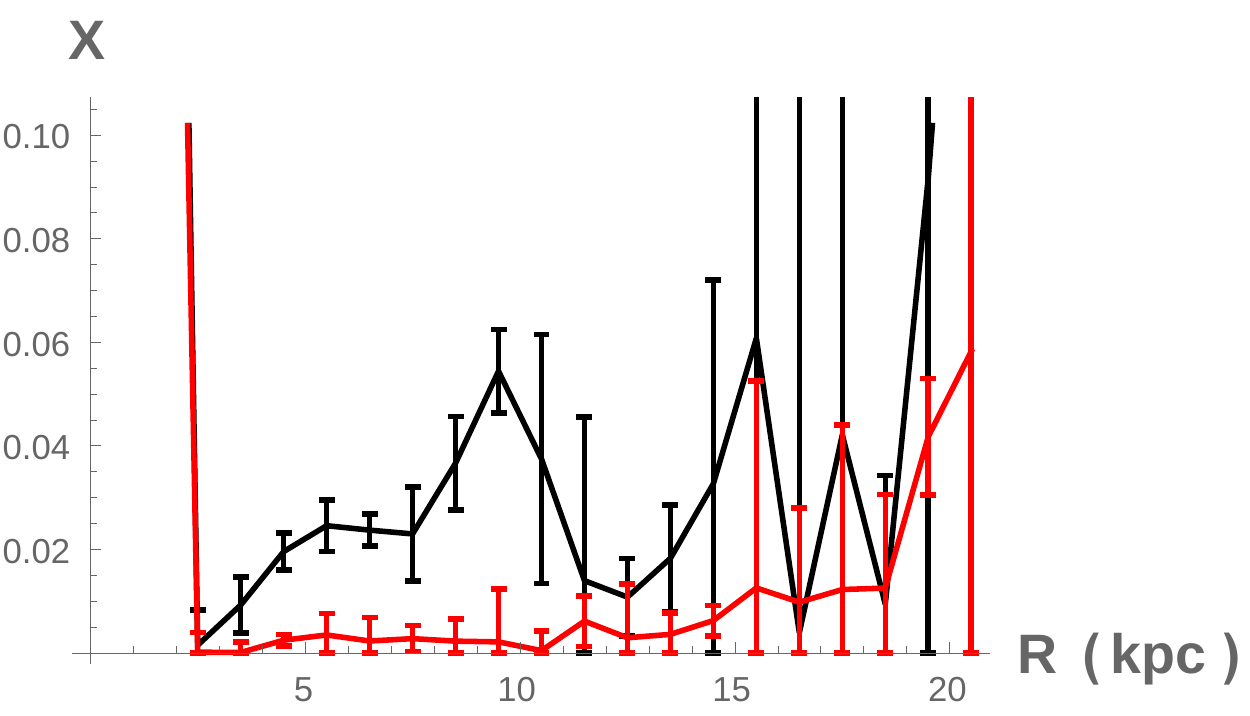}
\includegraphics[width=2.8in,height=1.6in]{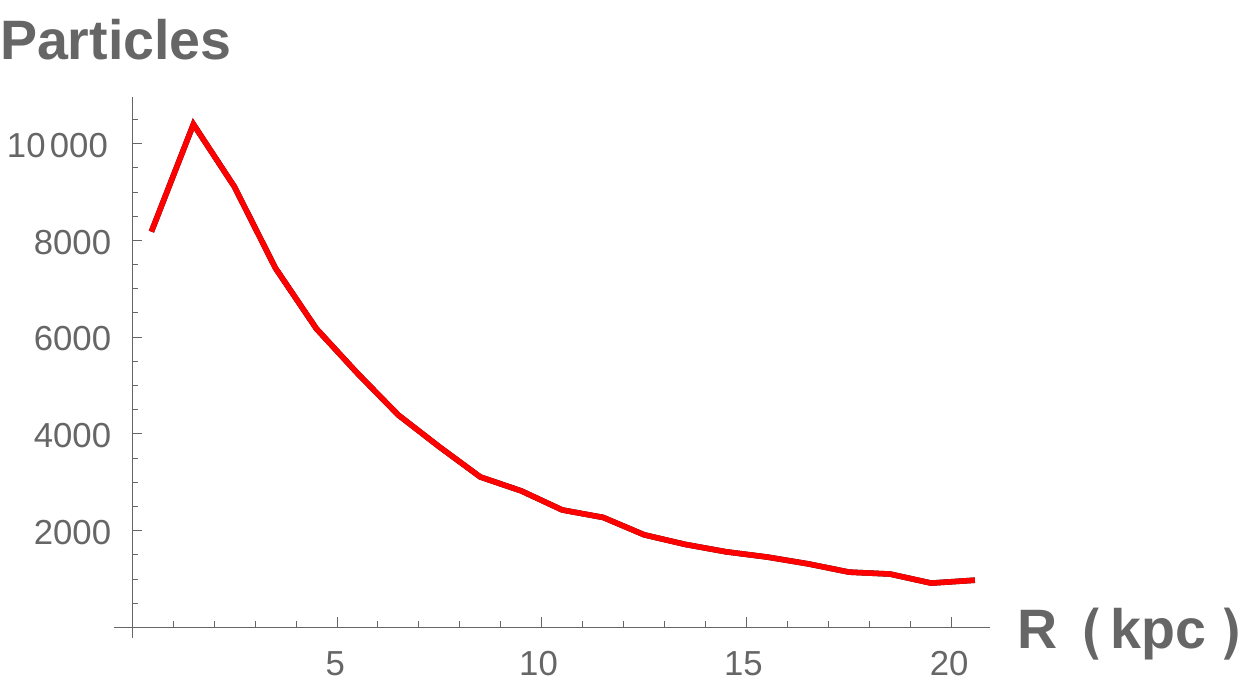}
\includegraphics[width=2.8in,height=1.6in]{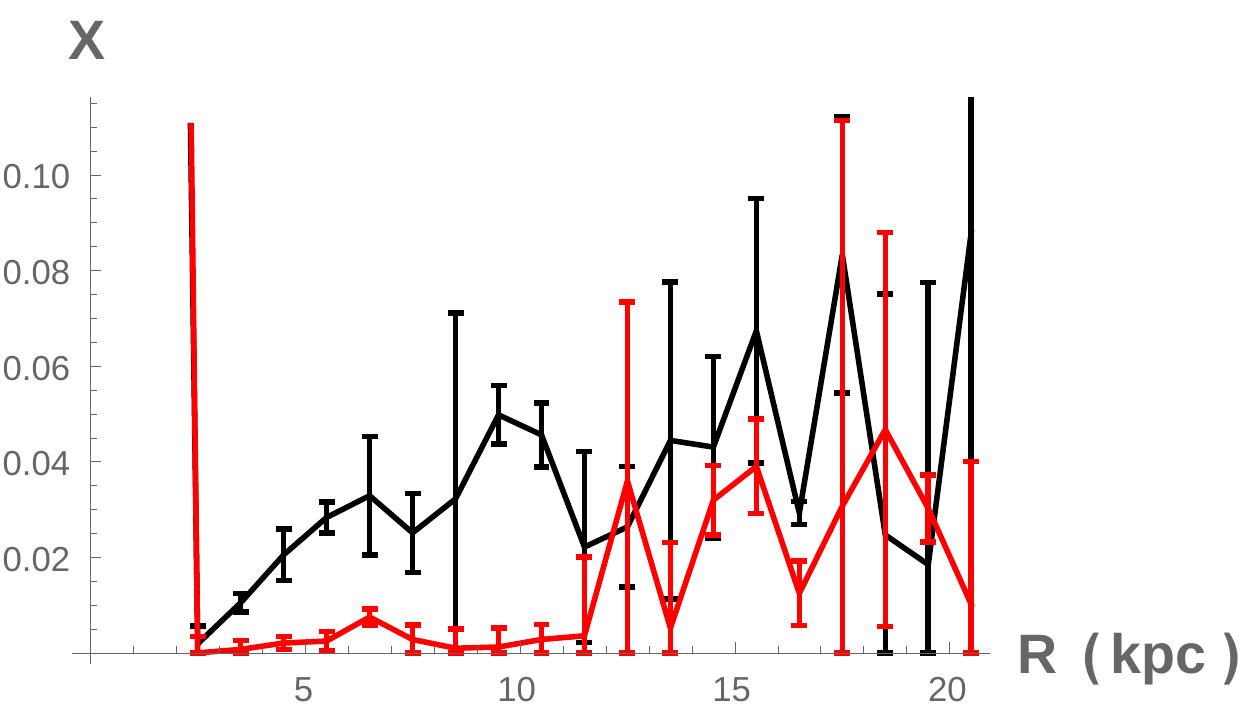}
\includegraphics[width=2.8in,height=1.6in]{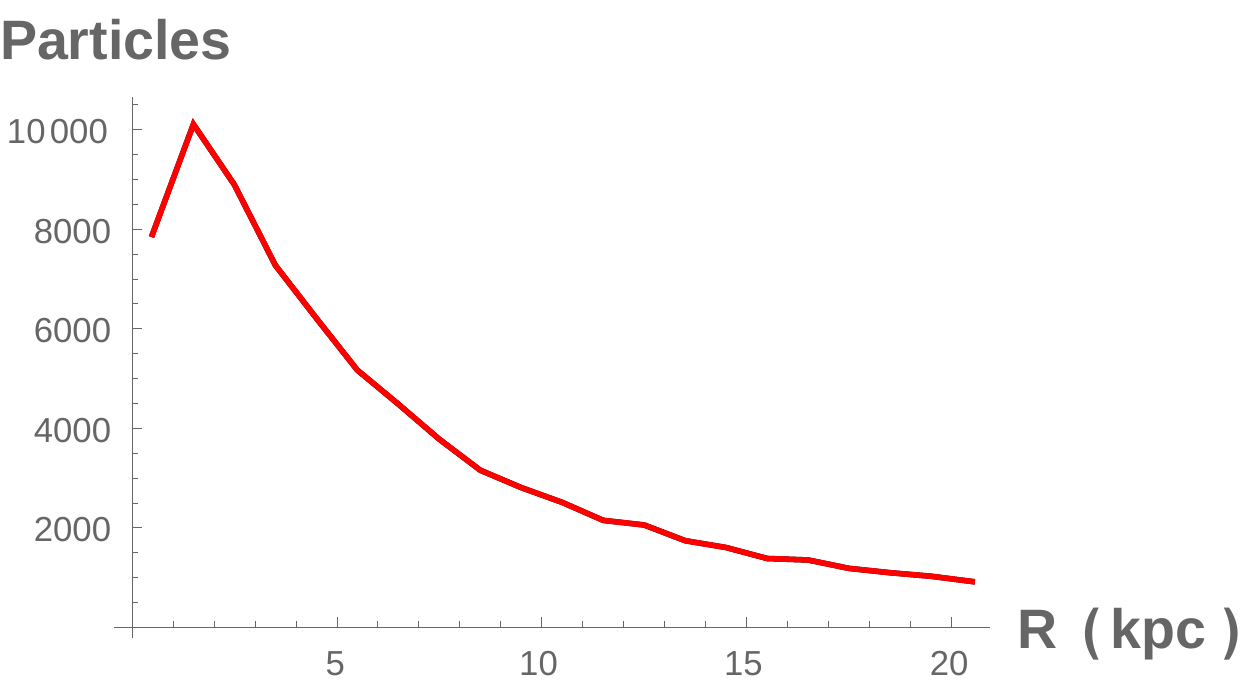}
\caption{The same three GALIC configurations with no tracer condition as are considered as in Fig.~\ref{notracefig}.  Only the line of sight projected information is used, where the line of sight is perpendicular to the symmetry axis of the halo.  These projected moments are deprojected using the inverse Abel transforms given in Sec.~\ref{projsez}.  Again each bin is 4 kpc.}
\label{projnotracefig}
\end{center}
\end{figure}

In Figs.~\ref{projfig} and \ref{projnotracefig} we present the equilibrium indicator $X$ obtained using only the line of sight projected moments, restricting and not restricting our attention to the tracer subpopulation respectively.  The main result of our analysis is that in general $X$, {\bbf{within the reported errors, is similar to or}} somewhat smaller using the projected moments.  The cause of this somewhat surprising result is that the inverse Abel transforms, which are used to deproject the data, enforce spherical symmetry, even when the system is not spherically symmetric. This enforcement has little effect on the second moments.  However the GALIC code begins by setting the second moments to their equilibrium values, it is the higher moments which evolve towards equilibrium in GALIC.  And the deprojection has an averaging effect on the fourth moments which in these cases drive them closer to an equilibrium configuration, even in the case of a spherically symmetric configuration.  {\bbf{This is because, when determining the moments from Eq.~(\ref{quadabel}), one imposes all of the relations between the fourth moments which would be implied by spherical symmetry, such as Eq.~(\ref{ideq}).  In our numerical calculations we have found that enforcing these relations greatly reduces the shot noise in $X$.  Intuitively this is quite reasonable: statistical fluctuations generally do not respect the spherical symmetry, and so by spherically averaging these fluctuations are reduced.}} 

How is it possible that with less information we obtained a better value of $X$?  The reason is that $X$ has simply become less sensitive to the details of the configuration, for example obviously $X$ is insensitive to a distortion along the line of sight direction.   Equilibrium implies $X=0$, but the converse is not true and in the projected cases the equilibrium test provided by $X$ is somewhat less powerful.

\section{Applications to Future Observations} \label{tmtsez}

To our knowledge, today there exists no astrophysical dataset to which these formulae can be usefully applied.  The problem is that the nearest collisionless, dispersion-supported systems, the Milky Way's satellite galaxies, are too far away for the proper motions of individual stars to have been determined with a precision of order the velocity dispersion.  However this will soon change.

{\bbf{In the next two or three years, the Gaia Space Telescope will determine the proper motions of more than 100 members of, for example, the Sculptor dwarf galaxy \citep{megaia}.  The precision, better than 20 km/sec will be comparable to, although somewhat greater than, the stellar dispersion, and so will already provide interesting hints of a core or cusp, especially when combining the results for multiple satellite galaxies.  However, in the present note we considered simulations with more than $10^4$ tracers and even so we saw that the determination of equilibrium is challenging.  Therefore with $10^2$ tracers Gaia will only be sensitive to extremely large departures from equilibrium.}}

{\bbf{However Gaia will not provide only proper motions, more importantly it has provided and is continuing to provide precise astrometric positions for hundreds of members of the Sculptor dwarf and other dSphs.  These positions may be combined with a single epoch of precise astrometry using  extremely large telescopes 10 years from now.  The long baseline between these observations will yield 5 km/sec precision proper motions for the Gaia sample, with 10 km/sec for hundreds of other Sculptor members which were also observed by Gaia.  At least in the spherically-symmetric case, this is already sufficient for a robust distinction of a core or cusp in a classical dSph if equilibrium can be shown to sufficient precision.}}

With a second epoch of observations, 5-7 years later, the Thirty Meter Telescope (TMT) will measure, with 3-5 km/sec precision, proper motions of $10^4$ or more stars in the largest classical dSphs.  As site selection for TMT is again open, with candidate sites in the northern and southern hemispheres, it is not yet clear which targets will be chosen.  However the results of this note suggest that one needs of order $10^4$ stars.  To obtain the proper motion of a dSph member at 80 kpc with an uncertainty beneath the stellar dispersion, given less than a minute of observations per field per epoch, one requires an $H$ magnitude of at most 22 \citep{tmt}.  Therefore only the classical dSphs are suitable targets for such a study.  Furthermore the Sculptor dwarf is quite large in the sky, and so would require several weeks of observing time per epoch.  While Fornax's greater distance implies greater uncertainties per star with a fixed viewing time, given the greater stellar population it remains a promising target if a southern hemisphere site is chosen.  The E-ELT is also capable of such measurements, but the systematics-limited astrometric precision will be 50 $\mu$as, as compared with 10-20 $\mu$as for TMT's first light IRIS instrument.  This means that E-ELT will in general measure proper motions with errors comparable to the intrinsic velocity dispersions.  {\bbbf{This is critical, because if the measurement errors are smaller than the intrinsic velocity dispersions, then they will provide subdominant contributions to the total error budget in the determination of quantities such as the halo shape and also the equilibrium parameter $X$.  Therefore we expect the above analysis of the uncertainties expected in $X$ to be applicable to these observations.}}

{\bbf{How might the results of this paper be applied to these datasets in practice?  Our main result is that the matrices $\alpha_{ij}$ and $\beta_{\ij}$, which may be summarised into $X_j$, characterize the departure from equilibrium.  More precisely, for a precisely spherical system with infinitely many observed tracers, $X_j=0$ in each radial bin $j$.  In practice, $X_j$ will be nonzero even for an equilibrium system as a result of shot noise, asphericity and measurement uncertainties.  In this paper we have studied the effects of shot noise and asphericity on $X$, while we believe that measurement uncertainties are unlikely to dominate over shot noise so long as these uncertainties are below the stellar dispersion, as is expected for the TMT observations described above.   Our studies of these sources of error used the GALIC code, which produces dark matter halos which are close to Hernquist profiles and a rather simple tracer condition.  To determine the uncertainties which are really expected in $X$, one needs to use a more realistic mass model for the dark matter and tracers, which of course needs to be determined by the data itself.}}

{\bbf{One way to implement this requirement is as follows.  First, one may assume equilibrium and analyse the data using a Jeans analysis or orbit based methods to derive a 3-dimensional mass distribution for the dark matter and the stars in a dSph.  Next, one can repeat an analysis like that done in this note to calculate the range of $X_j$ expected in an equilibrium configuration with such a distribution.  The observed $X_j$ can be compared with this range to determine the confidence with which the equilibrium hypothesis may be excluded.  This strategy may be applied to all of the observed stars, or to individual subpopulations.}}

\section{Conclusions} \label{consez}

We have provided a proof-in-principle that, for a spherically symmetric system, one may test equilibrium using the fourth order Jeans equations.  We have introduced a test statistic $X$ which tests the equilibrium hypothesis, such that for a spherically symmetric, equilibrium system $X$ tends to zero as the number of particles observed tends to infinity.  This property of $X$ requires no assumptions on the form of the radial-dependence of the gravitational potential, nor on the velocity anisotropy.

We have calculated $X$, bin by bin, in spherically symmetric and axisymmetric configurations.  We have considered both near-equilibrium configurations generated by GALIC and also samples drawn from exact equilibrium configurations, and their deformations.  We have analyzed these directly from the full 6-dimensional phase space and also using only the projected observables together with inverse Abel transforms.  In each case, we considered a sample of $10^5$ particles of which $1.8\times 10^4$ are tracers, with a radial and velocity distribution similar to stars in a very large dSph\footnote{{\bbf{Scale invariance of the equations of motion and the definition of $X$ implies that all of our results can be trivially scaled down to smaller dSphs.}}}.  Due to our discrete derivatives and other binning effects, our uncertainties were large in the inner 800 pc.  However, with this sample we were able to show that the indicator $X$ works quite well, with a precision and stability below $10^{-2}$ in the next 1 or 2 kpc.  At larger distances, the small number of stars begins to lead to large statistical fluctuations.

Although less information is available, we found that the same configurations lead to similar or smaller values of $X$ when only the projected data is used.  However in this case $X$ provides a less powerful test of the equilibrium hypothesis, as there are more nonequilibrium perturbations to which it is insensitive.

To actually conclude the significance with which equilibrium can be excluded, {\bbf{one may for example follow the procedure described at the end of Sec.~\ref{tmtsez}.}}  This can be {\bbf{done}} rather straightforwardly.  However {\bbf{one must recall that this procedure}} is suboptimal, {\bbf{in the sense that it does not deliver the smallest possible uncertainties.  This is because, as was described in Sec.~\ref{projsez}, our formulas for the deprojection did not use the full available data in the observed moments.   They were chosen because they are analytic formulas which are easily computed.  However a more precise deprojection could be obtained by selecting the 3-dimensional distribution whose projection best matches the observed projected moments, as characterized for example by a $\chi^2$ fit.  More specifically,}} one may parametrize the various spherical-coordinate functions and use (\ref{dueeq}) and (\ref{quattroeq}) to determine the corresponding projected moments.  Then a $\chi^2$, statistic can be used to find the best fit moments and their associated uncertainties.  These can be inserted into Eqs.~(\ref{maina}) and (\ref{mainb}) to test the equilibrium of the system, with the uncertainties in the fixed time spherical coordinate moments propagated in the usual way through these two equations.  

A quick and dirty approach to testing the equilibrium hypothesis is as follows.  First, note that to determine the spherical second moments it suffices to use two of the three equations in (\ref{dueeq}) and three of the six in (\ref{quattroeq}).  Therefore there are 60 distinct ways to determine the moments, which using (\ref{maina}) and (\ref{mainb}) yield 60 distinct estimates for the time derivatives.  {\bbf{One can then estimate the uncertainty in $X_j$ to simply be the square root of the variance of the 60 values of $X_j$ given by these 60 permutations of our procedure.  This source of error should be added in quadrature to other sources arising from shot noise and finite binning.  Then one may test the null hypothesis of equilibrium by simply calculating the number of standard deviations between $X$ and $0$.   This yields a very rough estimate of the significance with which the equilibrium hypothesis may be rejected.}}

To exclude equilibrium one needs a systematic investigation of the bias caused by the aspherical halo shape, for example via an analysis similar to that done in \citet{triax} or \citet{milking}.  The main shortcoming with our approach in its current form is the assumption of spherical symmetry.  {\bbf{The goal of our research program is to eventually generalize these results to aspherical systems.  While we have seen that our spherical results may be applied to an aspherical system without an enormous increase in the uncertainties, we believe that a more precise incorporation of the asphericity will strengthen our test of equilibrium.  To do this one will need to consider not only the radial dependence of the projected moments, but also the azimuthal dependence of these moments, as in \citet{meell}.}}

\section* {Acknowledgement}

\noindent
JE is supported by NSFC MianShang grant 11375201 and the Recruitment Program of High-end Foreign Experts and the CAS Key Research Program of Frontier Sciences grant QYZDY-SSW-SLH006.

\software{GALIC \citep{galic}
          }




\end{document}